\documentclass{article}

\usepackage{arxiv}

\usepackage[utf8]{inputenc} 
\usepackage[T1]{fontenc}    
\usepackage{hyperref}       
\usepackage{url}            
\usepackage{booktabs}       
\usepackage{amsfonts}       
\usepackage{nicefrac}       
\usepackage{microtype}      
\usepackage{lipsum}		
\usepackage{graphicx}
\usepackage[inline]{enumitem}
\usepackage{subfig}  
\usepackage{multicol}
\usepackage{multirow}
\usepackage{tabularx}
\usepackage{amsmath,amssymb}
\usepackage{color}
\usepackage{xspace}
\usepackage[table,xcdraw]{xcolor}
\usepackage[normalem]{ulem}
\usepackage{bm}
\usepackage{hyperref}
\usepackage{listings}
\usepackage{algorithm}
\usepackage{algpseudocode}
\usepackage{paralist}
\usepackage[export]{adjustbox}

\makeatletter
\DeclareRobustCommand\onedot{\futurelet\@let@token\@onedot}
\def\@onedot{\ifx\@let@token.\else.\null\fi\xspace}
\def\etal{~et~al\onedot}
\def\eg{e.g\onedot} 
\def\ie{i.e\onedot}

\makeatother

\def\clap#1{\hbox to 0pt{\hss #1\hss}}%
\def\initials#1{\protect\clap{\smash{\raisebox{1.4ex}{\tiny{\textsf{\textit{#1}}}}}}}%
\makeatletter
\newcommand{\EDIT}[4][]{\protect\@ifundefined{hidecomments}{%
  \strut{\color{#3}{\hspace{0pt}\initials{#2}\protect\sout{#1}{~#4}}}%
  }{}}
\newcommand{\NOTEboxed}[3]{\protect\@ifundefined{hidecomments}{%
  {\begin{center}\fbox{\parbox{0.97\linewidth}{\protect\EDIT{#1}{#2}{#3}}}\end{center}}
  }{}}
\makeatletter
\newcommand{\DefAuthor}[2] 
{%
  \expandafter\newcommand\csname #1edit\endcsname[2][]{\protect\EDIT[##1]{#1}{#2}{##2}}
  \expandafter\newcommand\csname #1\endcsname[1]{\protect\csname #1edit\endcsname{[##1]}}
  \expandafter\newcommand\csname #1boxed\endcsname[1]{\NOTEboxed{#1}{#2}{##1}}
}
\makeatother

\newcommand{\abs}[1]{ \left| #1 \right| }

\graphicspath{{Image/}{./}}

\newcommand{\myref}[2]{{\hyperref[#1]{\protect\ref{#1}#2}}}

\title{Glyph from Icon -- Automated Generation of Metaphoric Glyphs}

\author{
	Dmitri Presnov  \\
	Computer Graphics Group \\
	University of Siegen \\
	Siegen 57068, Germany\\
	\texttt{dmitri.presnov@uni-siegen.de}\\	
	\And
	Andreas Kolb \\
	Computer Graphics Group \\
	University of Siegen \\
	Siegen 57068, Germany\\
	\texttt{andreas.kolb@uni-siegen.de} \\
}

\begin{document}
\maketitle

\begin{abstract}
	Metaphoric glyphs enhance the readability and learnability of abstract glyphs used for the visualization of quantitative multidimensional data by building upon graphical entities that are intuitively related to the underlying problem domain. Their construction is, however, a predominantly manual process.
	In this paper, we introduce the \emph{Glyph-from-Icon (GfI)} approach that allows the automated generation of metaphoric glyphs from user specified icons.
	Our approach modifies the icon's visual appearance using up to seven quantifiable visual variables, three of which manipulate its geometry while four affect its color. Depending on the visualization goal, specific combinations of these visual variables define the glyphs's variables used for data encoding.
	Technically, we propose a diffusion-curve based parametric icon representation, which comprises the degrees-of-freedom related to the geometric and color-based visual variables.  Moreover, we extend our GfI approach to achieve scalability of the generated glyphs.
	Based on a user study we evaluate the perception of the glyph's main variables, \ie, amplitude and frequency of geometric and color modulation, as function of the stimuli and deduce functional relations as well as quantization levels to achieve perceptual monotonicity and readability.
	Finally, we propose a robustly perceivable combination of visual variables, which we apply to the visualization of COVID-19 data.
	
\end{abstract}

\begin{figure}[h!]
	\centering
	\includegraphics[width=\textwidth]{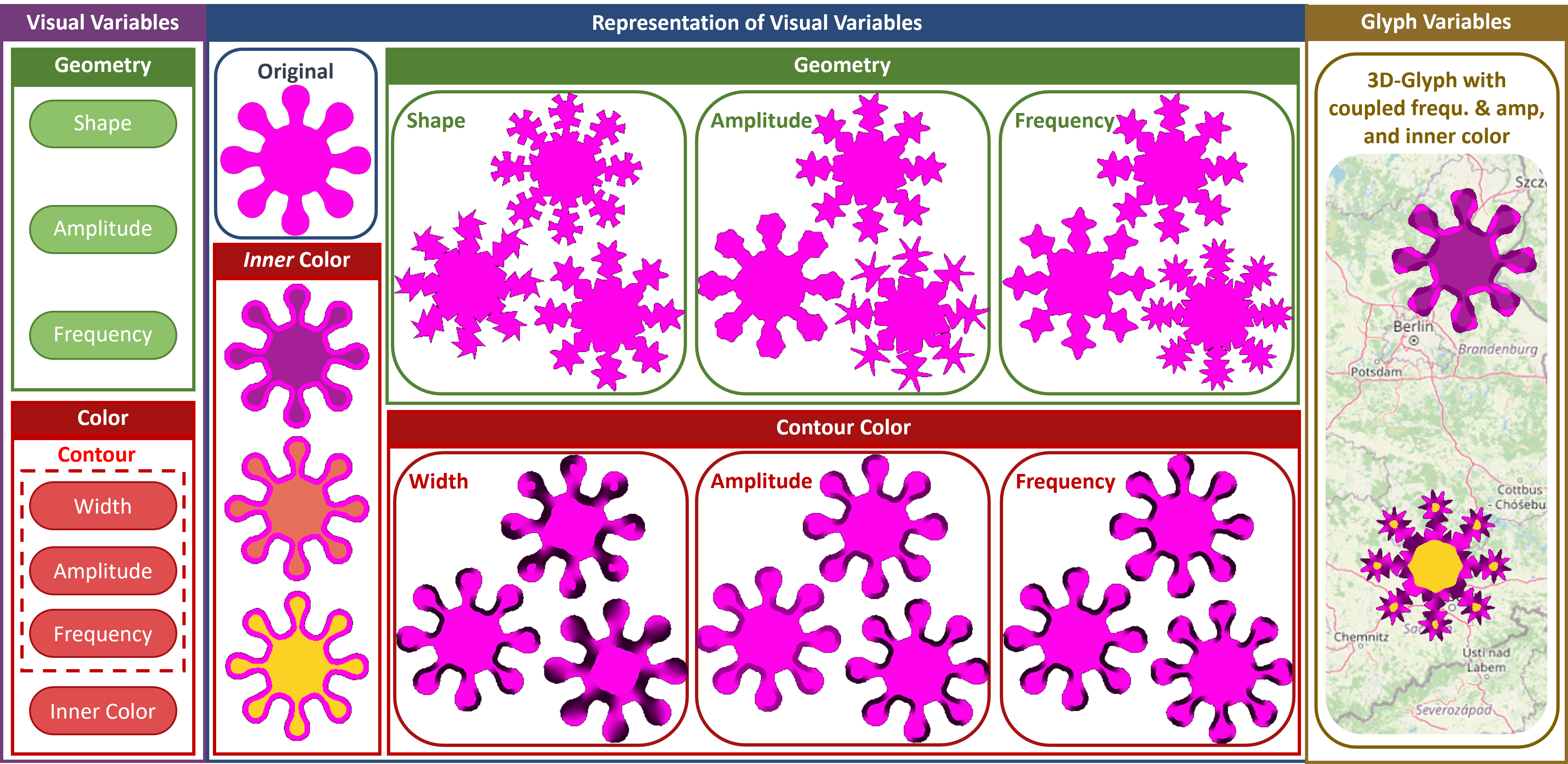} 
	\caption{The \emph{Glyph-from-Icon (GfI)} concept. The seven visual variables (dark purple box, left) are represented using extended parametric diffusion curve images that allow the automatically modification of an original icon (blue box, center). The final glyph can use several of the visual variables in parallel, and may combine them to enhance the readability of the quantitative values. The example in the  golden box on the right shows a ``3D'' glyph comprising three variables, \ie the combined geometric and color amplitude, the combined geometric and color frequency and the inner color.}
	\label{fig:concept}	
\end{figure}


	\section{Introduction}
	
	In visualization, (data) glyphs, developed from the 1950s onward, are used to represent and visualize multidimensional data by assigning data values to visual variables, such as size, color hue, luminance (or color value), grain, orientation, and shape~\cite{bertin1983semiology}, of a pre-defined graphical entity~\cite{fuchs2016systematic}. 	
	The primary goals of mapping data values to visual variables in glyph design are
	\emph{perceptual monotonicity}, \ie increases in data values result in corresponding perceptually increasing visual variables,
	\emph{accurate readability}, that is the ability to read data values from the displayed visual variables, and
	\emph{perceptual independence}, \ie the variation in one visual variable does not interfere with the readability of another one~\cite{ware2009quantitative}.
	The majority of glyphs in information visualization are \emph{abstract}, \ie they make use or are composed of basic geometric primitives such as lines, boxes or circular segments to encode quantitative multidimensional data.
	Abstract glyphs can, for instance, comprise several polygons with varying color, height, density, and regularity~\cite{healey1999large}, or more complex geometries such as stick figures or star glyphs~\cite{peng2004clutter}.
	%
	Contrary to abstract glyphs, \emph{metaphoric glyphs} enable the readability of multidimensional data by using familiar and well-understood visuo-spatial phenomena that often relate to the underlying problem domain~\cite{risch2008role}. 
	Chernoff faces~\cite{chernoff1973use} are a very famous example, which are well suited for the recognition of human related information~\cite{jacob1978facial}.
	Metaphoric glyphs have also been used, for instance, to visualize the health state of corn cobs~\cite{nocke2005iconbased} and  environmental data related, \eg, to forest fires\cite{fuchs2015glyph}.
	The main advantages of metaphoric glyphs relate to their potential of increased readability in case of realistic glyphs~\protect\cite{flury1981graphical}, improved data understanding~\protect\cite{fuchs2016systematic}, \eg, by mapping  data to corresponding glyph parts~\protect\cite{surtola2005effect}.
	
	However, there is comparably little research work regarding the  (semi-)automated generation of metaphoric glyphs and the evaluation of the limits of their perceptual monotonicity, accurate readability and perceptual independence.
	One line of research strives for data driven and interactive methods for generating metaphoric glyphs, such as the automatic generation of emoji-like metaphoric glyphs~\cite{cunha2018many},
	which mainly focus on the variation of visual variables without evaluating quantitative monotonicity, readability and independence.
	A second line of research circumvents the problem of generating readable metaphoric icons by coupling metaphoric and abstract glyph components in a \emph{loose fashion}, that is by putting them side-by-side~\cite{legg2012matchpad, chung2013sorting}.
	In this paper, we introduce the \emph{Glyph-from-Icon (GfI)} approach, a novel technique that allows the automated generation of metaphoric glyphs from icons in a \emph{tight fashion} by directly modifying their shape and color appearance.
	We utilize \emph{diffusion curves}~\cite{orzan2008diffusion} as parametric representation of a user given icon, which we reparameterize to add new degrees of freedom (DOF) in an arc-length encoded manner. Using these DOFs allows us to modify the shape and color appearance in a visual consistent way. That is, we aim at retaining the icon's overall shape and hue, as they are related to the application domain.
	The GfI concept presented in this paper comprises seven visual variables: wave-like geometric modification of the icon contours (shape, amplitude and frequency), periodic modification of the contour luminance (amplitude, frequency, contour region width) and inner color.
	We perform a user study to calibrate the mapping between stimuli and perceived visual variables to achieve optimal perceptual monotonicity, accurate readability and perceptual independence.
	In summary, our paper comprises the following contributions.
	\begin{compactitem}
		\item Diffusion-curve based \emph{Glyph-from-Icon (GfI)}, a new technique to automatically generate glyphs as modifications of icons. The modification process comprises
		\begin{compactitem}
			\item the reparameterization of the icon contours as diffusion curves using B-splines and the methodology to insert the DOFs in an arc-length encoded manner required to encode the visual variables,
			\item the wave-like modification of the curves, preventing self-intersection using distance transforms, and
			\item the modification of the icon contours' color parameters in a periodic fashion in combination with the automated subdivision of the icon into a contour and an inner regions.
		\end{compactitem}
		\item The quantitative survey of GfIs to achieve perceptual monotonicity and independence, and to derive proper quantization to achieve accurate readability.
		\item We present several application examples of GfIs.
	\end{compactitem}
	
	\section{Prior Work}
	\label{s:prior}
	
	We briefly discuss prior work conceptually related to our curve-based Glyph-from-Icon (GfI) concept, \ie, metaphoric glyphs, contour-based approaches to quantitative (uncertainty) visualization, and diffusion curves.
	
	\paragraph*{Metaphoric Glyphs.}
	Metaphoric glyphs form a specific sub-group of glyphs that try to enhance the underlying communication process by additionally utilizing visual analogies from the related application domain, and ultimately strive for ``the picture becomes the thing it represents''~\cite{risch2008role}. 
	Several works underline their potential to improve readability~\cite{fuchs2016systematic}. Example studies are the comparison between abstract face glyphs and metaphoric car glyphs for car-related data using, \eg, horsepower to the size of a car's hood~\cite{surtola2005effect}, 
	and between metaphoric RoseShape glyphs~\cite{cai2015applying} and abstract regular n-sided polygonal glyphs for multidimensional educational data~\cite{li2015metaphoric}.
	For a general overview of glyph design and application, we refer the reader to the surveys from Ward~\cite{ward2008multivariate}, Borgo\etal\cite{borgo2013glyph} and Fuchs\etal\cite{fuchs2016systematic}. 
	
	Various task specific designs for metaphoric glyphs have been presented in the past.
	Nocke\etal\cite{nocke2005iconbased} propose a design method for metaphoric glyphs using a mosaic paradigm that decomposes the icon into tiles, alters the tiles in size, shape or color according to the data values, and recombines them to achieve the glyph.
	Fuchs\etal\cite{fuchs2015glyph} propose a manual design of leave-shaped glyphs to utilize the humans ability to visually discriminate natural shapes by modifying, \eg, the leaf's morphology, venation, and boundary shape for  visualizing multidimensional data related to environmental events such as forest fires.
	Based on the loose combination of metaphoric icons and abstract glyph components by Legg\etal\cite{legg2012matchpad}, Chung\etal\cite{chung2013sorting} propose wave-like shape deformations and color modifications to a circle to encode a sportsman's performance, which surrounding the icon that describes the specific sports event to be assessed.
	Cunha\etal\cite{cunha2018many} present a data-driven strategy for the automatic generation of emoji-like metaphoric glyphs that uses the structured emojinating database~\cite{cunha2018emojinating}. Their system suggests glyphs as variations of the database emojis for a specific theme.
	Recently, Ying\etal\cite{ying2021glyphcreator} presented GlyphCreator, a tool for deep learning based decomposition of circle-like abstract glyphs into several visual elements, which are then manually bound to the input data attributes.
	\emph{Summary.} Existing approaches to metaphoric glyph are largely dominated by manual design processes or structured databases. Furthermore, to our knowledge, while few works conduct expert interviews, \eg \cite{ying2021glyphcreator}, quantitative user studies have not been conducted for any of the (semi-)automated techniques for generating metaphorical glyphs.
	
	\paragraph*{Contour-based Methods.}
	Even though abstract glyphs are frequently used in visualization, we mainly focus on contour-based uncertainty visualization methods~\cite{bonneau2014overview}, as these approaches relate to the general question of how to modify icons in a controllable fashion.
	
	Various contour-based approaches have been applied to encode uncertainty in the underlying data. This includes varying the contour lines' width~\cite{allendes2008contouring}, using sets of contours in regions with high uncertainty in segmentation~\cite{prassni2010uncertainty}, or modifying  graph nodes and radial color gradients to encode the relation uncertainties in graphs.
	G\"ortler\etal\cite{gortler2017bubble} propose bubble treemaps as an extension of circular treemaps that encode uncertainty using wave-like modifications and blur-effects applied to the circular-arc spline contours during the treemap generation process. 
	Holliman\etal\cite{holliman2019visual} use abstract circle-shaped glyphs for uncertainty visualization with wave-like modified contours, manually modeled in blender.
	Both techniques are conceptually related to our Glyph-from-Icon concept in using wave-like contour modifications.

	\emph{Summary.} There exist several concepts to use shape and appearance modification in uncertainty visualization. In all approaches, the modifications are directly integrated in the construction process of \emph{pre-defined} shapes or abstract glyphs. Only recently, Holliman\etal\cite{holliman2019visual} perceptually evaluated some aspects regarding the effectiveness and readability of this visualization concepts.
	
	\paragraph*{Diffusion Curves}
	While approaches in contour-based (uncertainty) visualization rely on pre-defined shapes (see above), vector-based representations allow a more flexible handling of arbitrary shapes.
	Based on early work from Elder and Goldberg~\cite{elder1998image}, diffusion curves images (DCI) have been proposed by Orzan\etal\cite{orzan2008diffusion}. DCIs are a widely used vector graphics representation, for which various functionalities and improvements have been developed over the years. The main focus lies on the conversion of raster images to DCI~\cite{lu2019depth}, enhanced DCI representation~\cite{xie2014hierarchical}, DCI editing~\cite{jeschke2011estimating,jeschke2016generalized,lu2020shape}, and efficient rendering of DCIs~\cite{jeschke2009gpu,sun2014fast}.
	The editing and manipulation approaches proposed so-far for DCI are all based on manual intervention. Examples include the methods of Jeschke and colleagues~\cite{jeschke2016generalized,jeschke2011estimating}, who propose a click-and-drag metaphor for manipulating diffusion curve properties, and Lu\etal\cite{lu2020shape}, who present a combination of global and local deformations for manipulating coarse and fine image content, respectively.
	
	\emph{Summary.} While there exists a rich tool set for utilizing DCI, there is no adequate approach to control the diffusion curves appearance as required for glyph-like functionalities.
	%
	\section{The Overall Icon-to-Glyph Concept}
	\label{s:concept}
	The main motivation of our Glyph-from-Icon (GfI) approach is to combine the intuitiveness of icon-based visualization with the ability of glyphs to provide quantitative, multidimensional visualization. The underlying concept for an automated generation of glyphs comprises the following stages (see Fig.~\ref{fig:concept} for an illustration).
	\begin{compactenum}
		\item \emph{Definition of the visual variables.}
		In general, visual variables can be defined by variation of geometric, color and texture properties. Even though our approach works for arbitrary colored icons, we assume to have ``simple'' mono-colored icons as starting point for generating quantifiable visual variables by modifying the icon's geometry and color.
		\begin{compactenum}
			\item Modifying the \emph{geometry} refers to the variation of the icon's contours using periodic modulations. This relates to three visual variables, namely the \emph{shape} of the modulation, \eg sinusoidal, the \emph{geometry modulation frequency} and \emph{geometry modulation amplitude}, whereby the latter two are quantifiable.
			\item Modification of the \emph{color} is applied to two regions, the \emph{contour region} (or margin) and the \emph{inner region} of the icon. Similar to the contour geometry, we use periodic modification of the icon's color in the contour region. The resulting visual variables are the \emph{color modulation frequency} and the \emph{color modulation amplitude}, as well as \emph{margin width} and, finally, the icon's \emph{color in the inner region}.  All color visual variables are quantifiable.
		\end{compactenum}
		\item \emph{Representation of the visual variables.}
		Technically, we modify the visual variables by converting the given icon into a \emph{diffusion curve image (DCI)} representation, adding new DOFs, which provide a parametric control over the icons geometric and color properties, enhanced by functionalities for prevention of interferences, such as self-intersections of contours. The separation between the icon's contour and inner region is achieved by inserting diffusion barriers.
		
		All visual variables are automatically generated by algorithmically inserting and modifying DC parameters as discussed in Sec.~\ref{s:pipeline}.
		
		\item \emph{Definition of the final glyph variables.}
		When setting up a visual data encoding for the glyph, the following issues have to be taken into account. On the one hand, the visual variables are not completely independent from each other, \eg, the amplitude is visually ``undefined'' for a frequency of zero. On the other hand, the relation between the stimulus values and the perceived magnitudes is not linear, which also requires the investigation of an adequate step-size, similar to a just-noticeable-difference (JND).
		We performed a user study to determine these relations and step-sizes and, furthermore, derived rules for adaptation of the glyph variables and step-sizes for scaled glyphs. Moreover, it might be advisable to combine several visual variables to gain more robustness in the visual readability of the glyph variable values.
		
	\end{compactenum}
	
	Obviously, there are also limits in terms of the initial icon's geometric  and/or coloration complexity, \ie, generating glyphs for too complex icons can impair the perception of the visual variables used for encoding.

	\section{The Icon-to-Glyph Pipeline}
	\label{s:pipeline}
	\begin{figure}[tb]
		\centering
		\resizebox{\columnwidth}{!}{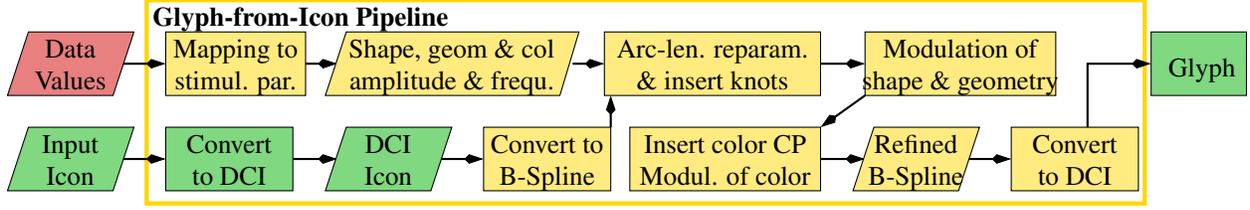}
		\caption{An overview of the Icon-to-Glyph Pipeline.}
		\label{fig:pipeline}	
	\end{figure}
	\begin{figure*}[tb]
		\centering
		\includegraphics[width=0.75\linewidth]{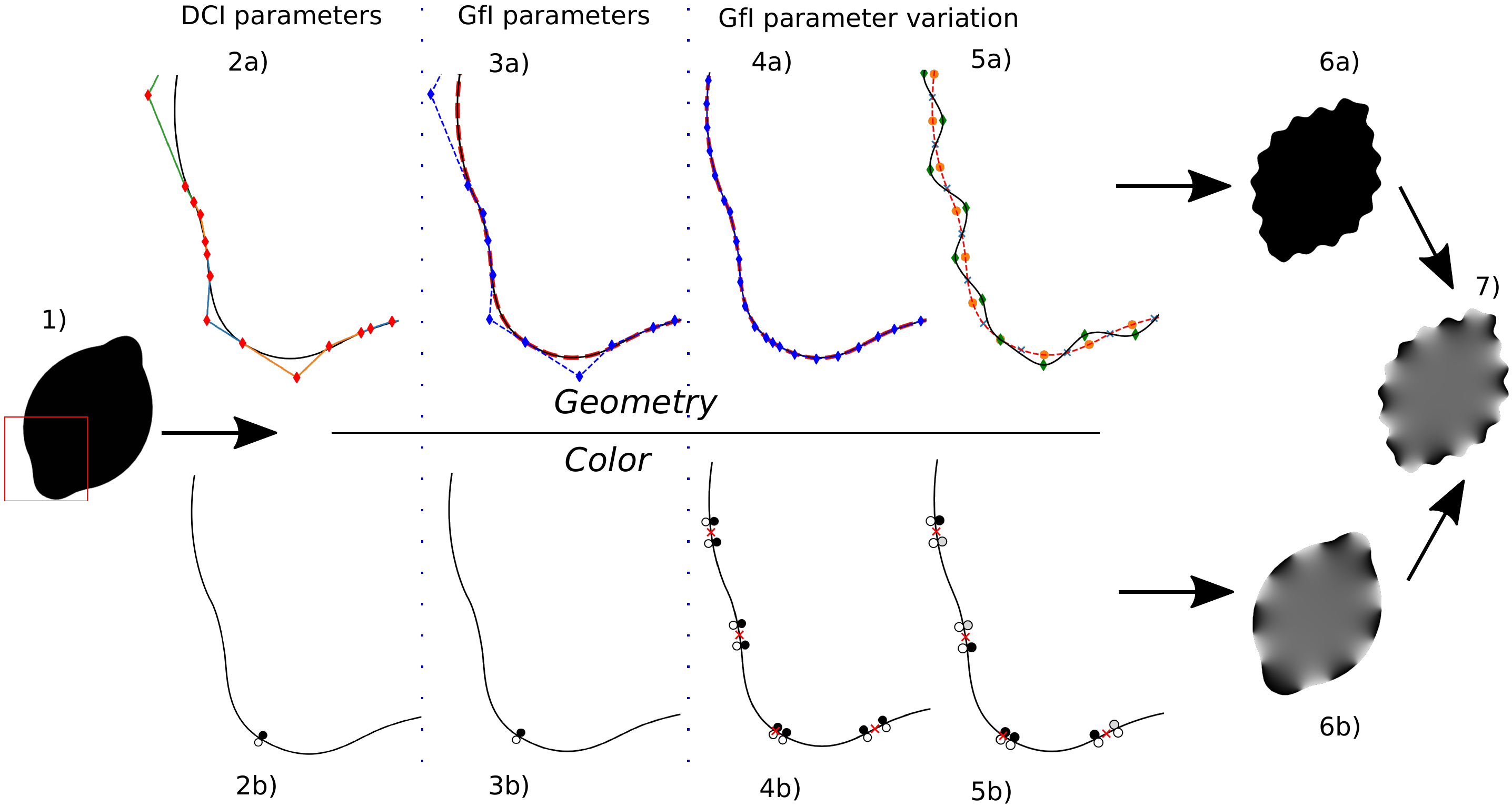} 
		\caption{The GfI pipeline with infinite contour width: The input raster image (1) is converted into a DCI, comprising B\'{e}zier control points for geometry (2a) and color control points (2b). Afterwards, the geometry is converted into B-splines and a $C^1$ approximation is applied (3a) w/o changing the color control points (3b). In the next step, DOFs are added according to the target geometry and color frequency, resulting in an arc-length reparametrization (4a) and new color points at the virtual borders of color intervals (4b: red crosses), respectively. The final shape modification transforms subgroups of knot points (5a: orange points are moved to green points; blue crosses remain unchanged). The color modification changes the color points of each second interval (5b: gray  circles) at the icon's inside, while the outer color is masked out and remains. The final glyph is obtained by re-conversion into standard DCI and rendering (6a, b; 7).}	
		\label{fig:params}	
	\end{figure*}
	This section describes the construction of \emph{Glyph-from-Icon (GfIs)} by augmenting a given icon with geometric and color visual variables (see Sec.~\ref{s:concept} and Fig.~\ref{fig:concept}). The core of this process is a controllable and automatic application of visual variables facilitated by a \emph{parametric representation} of icons. The GfI parametrization is based on DCIs~\cite{orzan2008diffusion}, which represent an image as a set of cubic B\'ezier splines in conjunction with color and blur attributes, which are tied to a parametric position $u$ along the respective spline (our current GfI approach omits the blur parameter due to its visual dependency to color attributes). The relevant DCI parameters are B\'ezier control points $\left\{\mathbf{P}_i=(x,y)\right\}_{i=0}^K$ for geometry and color control points $\left\{\mathbf{C}^l_i=(r,g,b,u)\right\}^L_{i=0}$, $\left\{\mathbf{C}^r_i=(r,g,b,u)\right\}^M_{i=0}$ for color on the left and right curve side, respectively (see also Fig.~\myref{fig:params}{:2a-b}). Proceeding from this data structure, GfI replaces the cubic B\'ezier splines with cubic B-spline curves and adjusts the parametric positions of the respective color attributes accordingly. Thus, the GfI parameters are de~Boor points $\left\{\mathbf{D}_i=(x,y)\right\}_{i=0}^N$ and a corresponding knot vector $\left\{t_i\right\}_{i=0}^{N+4}$ for geometry and $\left\{\mathbf{C'}^l_i=(r,g,b,u')\right\}^L_{i=0}$, $\left\{\mathbf{C'}^r_i=(r,g,b,u')\right\}^M_{i=0}$ for color (see also Fig.~\myref{fig:params}{:3a-b}).
	
	The overall GfI constructing process is depicted in Fig.~\ref{fig:pipeline} and consists of the following steps.
	\begin{compactdesc}
		\item[Input and Pre-Processing:] Given an input icon (Fig.~\myref{fig:params}{:1}), we first transform it into a DCI (Fig.~\myref{fig:params}{:2a-b}) and subsequently in a GfI (Fig.~\myref{fig:params}{:3a-b}) with a resulting set of control parameters as described above (see also Sec.~\ref{s:pipeline.preproc}). 
		\item[Visual Variables:] Having the GfI parameters, we vary them to apply visual variables. For geometry, the knot vectors are modified according to the target shape and frequency, comprising the arc-length parametrization and altering the knot multiplicity (Fig.~\myref{fig:params}{:4a}), and the positions of the de~Boor points according to the target amplitude (Fig.~\myref{fig:params}{:5a}).
		We control the icon geometry by modifying the curve's \emph{knot points}, \ie the curve points at knot parameters $u=t_i$, and implicitly adapt the de~Boor points using energy minimization (see Sec.~\ref{s:pipeline.var.geometry}).  For color, the parametric positions of color attributes are set according to the target color frequency (Fig.~\myref{fig:params}{:4b}) and the respective RGB values are modified according to the target color amplitude (Fig.~\myref{fig:params}{:5b}). The width of the contour region is set by adding another diffusion curve, \ie a \emph{diffusion barrier} (see~\cite{Bezerra2010DiffBarriers}), which is defined as an isoline of the given distance from the icon's contour (see Sec.~\ref{s:pipeline.var.color}). 
		\item[Post-Processing:] The modified GfI representation is back-transformed into an equivalent DCI, which includes a transformation of B-splines into B\'ezier splines and a corresponding re-mapping of the parametric positions of color attributes (see Sec.~\ref{s:pipeline.postproc}).
	\end{compactdesc}
	
	\subsection{Input and Pre-Processing}
	\label{s:pipeline.preproc}
	We use the diffusion curves drawing tool from Orzan\etal\cite{orzan2008diffusion} to convert the given raster icon into a DCI. 
	As our contour modification requires as high continuity as possible to prevent unwanted cracks, we convert the cubic B\'{e}zier splines from DCIs into cubic B-spline curves, whereby the $C^0$ transitions between B\'{e}zier spline segments are approximated by $C^1$-transitions, in the cases where the first derivatives on both sides have (approximately) the same direction and differ only in length, as this can be corrected by an appropriate interval scaling. Thus, the multiplicity of the respective 3-folded knots can be reduced to 2 (see Fig.~\myref{fig:params}{:3a}).

	\begin{algorithm}[bt]
		\caption{Conversion of a B\'{e}zier spline into a B-spline curve.}
		\label{alg:bez2Bspline}
		\begin{algorithmic}
			\State $\mathbf{D} \leftarrow {\mathcal P}^{0}$
			\State $\mathbf{T} \leftarrow [0,0,0,0,1,1,1]$ 
			\State $\Delta \leftarrow 1$ \Comment{Length of first parameter interval}
			\For{$i\leftarrow 0$ to $N-2$} 
			\State $\Delta \leftarrow \Delta * \vert(b^{i+1})'(t_{i+1})\vert / \vert(b^{i})'(t_{i+1})\vert$ 
			\State $t_{next}\leftarrow \mathbf{T}.\mathtt{last()}+\Delta$
			\If{$\angle ((b^{i})'(t_{i+1})$, $(b^{i+1})'(t_{i+1})) < \alpha$}	
			\State $\mathbf{T}.\mathtt{append(}[t_{next}, t_{next}]\mathtt{)}$  \Comment{approx. $C^1$-continuous}
			\State $\mathbf{D}.\mathtt{pop()}$ \Comment{Delete the last control point}
			\Else \Comment{$C^0$-continuous}
			\State $\mathbf{T}.\mathtt{append(}[t_{next}, t_{next}, t_{next}]\mathtt{)}$
			\EndIf
			\State $\mathbf{D}.\mathtt{append(}{\mathcal P}^{i+1}\mathtt{)}$  
			\EndFor
			\State $\mathbf{T}.\mathtt{append(}\mathbf{T}.\mathtt{last())}$\Comment{4-fold knot at the B-spline's end}
		\end{algorithmic}
	\end{algorithm}
	Technically, the conversion into B-splines proceeds as follows (see Alg.~\ref{alg:bez2Bspline}). Given a B\'{e}zier spline with $N$ segments $b^l(u),\;i=0,\ldots,N-1$ with control points ${\mathcal P}^l=\left\{\mathbf{P}_{3l},\cdots , \mathbf{P}_{3l+3}\right\}$. We simply convert the $l$-th B\'{e}zier spline segment by adapting the parameter interval to fit the tangent length of the end point of the prior segment. In case the directional deviation is below the threshold $\alpha$ (we used $\alpha=3.5^{\circ}$), the transition is assumed to be $C^1$, and we append  a double knot to the knot-vector $\mathbf{T}$ and drop the last de~Boor point from the control point list $\mathbf{D}$.
	Subsequently, the parametric positions of color attributes are transformed by mapping them onto the new knot vector $\mathbf{T}$.

	\subsection{Applying Visual Variables}
	\label{s:pipeline.var}
	Initially, let us consider the construction of glyphs from icons by modifying its geometry and color without an inner region, \ie the margin width is infinite. Both modifications can be applied independently (Fig.~\myref{fig:params}{:6a-6b}) or in combination (Fig.~\myref{fig:params}{:7}).
	
	\subsubsection{Geometry}
	\label{s:pipeline.var.geometry}
	The wave-like modification of the curve geometry is performed by a periodic translation of the knot points, \ie the curve points for knot parameters, along the curve's normals in alternating directions as follows.
	
	\paragraph*{Frequency}
	is controlled via the period length of the wave, \ie 1/frequency. Depending on the requested shape, we add two knot points (see, \eg, de~Boor \cite{deBoor1978Guide}) to each period that will be translated orthogonal to the curve's tangent (`wave peaks'). This requires an appropriate knot vector of the B-spline, such that their corresponding knot points are placed equidistant on the curve. We, therefore, numerically calculate the arc length of the curve and store it in a lookup table. In general, the B-splines created in the pre-processing step (see Sec.~\ref{s:pipeline.preproc}) have not enough DOFs to interpolate the translated curve points, therefore we insert further knots using Boehm's algorithm~\cite{Boehm1980Inserting}.  After the new knot vector has been defined, the respective de~Boor points are adjusted by means of the least-squares progressive iterative approximation algorithm with energy term (ELSPIA)~\cite{he2015chord}), which minimizes the least-square distance to the original curve taking into account a stretching term. The result is a cubic B-spline with single knots, whose knot points are approximately equidistant along the curve (Fig.~\myref{fig:params}{:4a}), and which approximates the original curve sufficiently accurate.  Depending on the required shape, additional knots may be added as described below.
	
	The entire arc-length parametrization process of a B-spline with initial knot vector $\mathbf{T}$ and de~Boor points $\mathbf{D}$ for a given frequency can be summarized as follows:
	\begin{compactenum}
		\item Compute the new knot vector $\mathbf{T}'$ as parameters corresponding to the equidistant positions along the curve according to the given period length.
		\item Using the original B-spline, expand the knot vector $\mathbf{T}$ to $\mathbf{T}^{tmp}, \abs{\mathbf{T}^{tmp}}= \abs{\mathbf{T'}}$, by means of Boehm's algorithm (details of this expansion see below). The resulting intermediate B-spline with de~Boor points $\mathbf{D}^{tmp}$ and knot vector $\mathbf{T}^{tmp}$ has the \emph{identical geometry} as $\left(\mathbf{T}, \mathbf{D}\right)$.
		\item Construct the new B-spline $\left(\mathbf{T'}, \mathbf{D'}\right)$ from the \emph{approximated} curve $\left(\mathbf{T}',\mathbf{D}^{tmp}\right)$ using ELSPIA.
	\end{compactenum}

	Both, the final approximation precision and the optimization speed in step~3 depend on the difference between the input curve $\left(\mathbf{T}', \mathbf{D}^{tmp}\right)$ and the reference curve $\left(\mathbf{T}, \mathbf{D}\right)=\left(\mathbf{T}^{tmp}, \mathbf{D}^{tmp}\right)$. Therefore, we minimize the distance between both curves by inserting those knots $t' \in\mathbf{T}'$ into $\mathbf{T}^{tmp}$, whose corresponding knot points have as large distances to the knot points of $\mathbf{T}$ as possible, so that the necessary shift to be applied to the de~Boor points $\mathbf{D}^{tmp}$ to get $\mathbf{D}'$ after exchanging $\mathbf{T}^{tmp}$ by $\mathbf{T}'$ is as small as possible.
	
\begin{figure}[bt]
		\centering
		\resizebox{.48\columnwidth}{!}{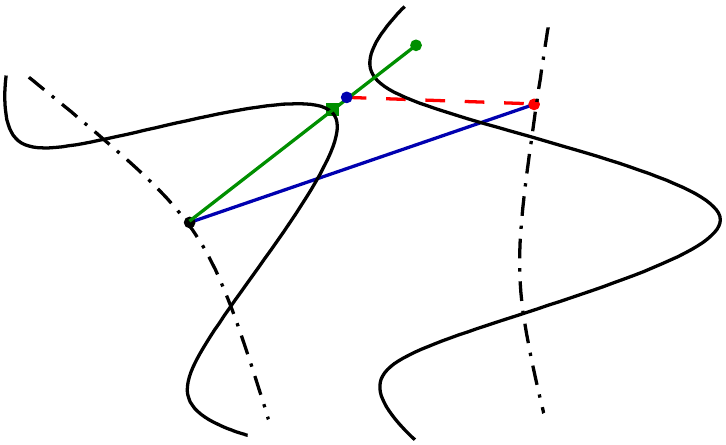}
		\includegraphics[width=0.48\columnwidth]{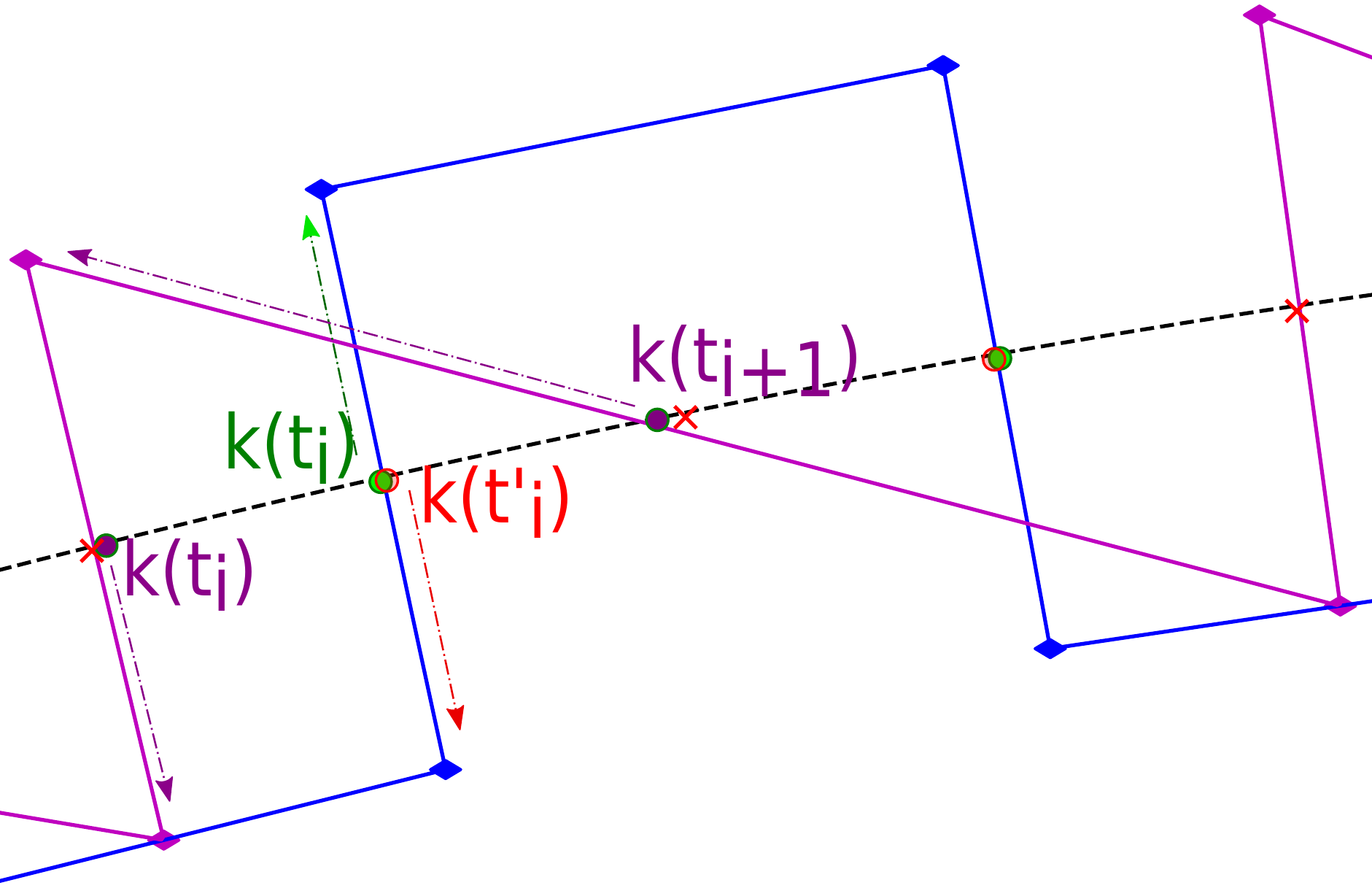} 
		\caption{Left: Prevention of curve intersection. Dashed lines show the original curves and solid lines their sinusoidal modification. The skeleton point $\mathbf{S}$ at the translation vector is equidistant to both curves. The sin peak (green square) has a pixel offset $\epsilon=3$ from $\mathbf{S}$.\newline
			Right: Construction of rectangular and sawtooth shapes of a B-Spline $k(u)$; arrows represent the respective translation vectors. Rectangular: additional knot points with a small offset (red) are added to  the initial knot point for the given frequency (green), and are translated in the opposite direction. Sawtooth: the target position for each second knot point is shifted.}
		\label{fig:skeleton-shapes}
	\end{figure}

	\paragraph*{Amplitude} is controlled by length of the translation vector applied to knot points that correspond to the intended shape (Figs.~\myref{fig:params}{:5a}, \ref{fig:skeleton-shapes}, right), which are used to constraint the modified B-spline $\left(\mathbf{T'}, \mathbf{D'}\right)$. We compute the de~Boor point offset vector $\Delta \mathbf{D}$ according to Fowler and Bartels~\cite{fowler1993constraint}:
	\begin{equation}
	\label{eq:splineMod}
	\Delta \mathbf{D}^T = \mathbf{B}^T (\mathbf{B} \mathbf{B}^T)^{-1}\Delta \mathbf{Q}^T\text{,}
	\end{equation}
	where $\Delta \mathbf{Q}$ is the offset vector to the translated knot points and $\mathbf{B}$ is the B-spline basis matrix.	
	
	Depending on the geometry of the original icon, the translation can lead to self-intersections that have to be prevented by locally reducing the amplitude, \ie the length of the translation vector. 
	
	The intersections prevention uses a subpixel precision Distance Transform (DT) of the original curves~\cite{maurer2003linear} to check if a translated knot point $\mathbf{K}+a\cdot\hat{\mathbf{n}}^K$ is closer to another curve point $\mathbf{N}$ than to original knot point $\mathbf{K}$ itself. In this case, we locally reconstruct a \emph{skeleton point} $\mathbf{S}= \mathbf{K}+a'\cdot\hat{\mathbf{n}}^B$ along the translation vector that is equidistant to the $\mathbf{K}$ and $\mathbf{N}$, whose amplitude $a'$ is given as the length of a leg in the isosceles triangle with base $\mathbf{N} - \mathbf{K}$, \ie $a' = \frac{0.5 \lVert \mathbf{N} - \mathbf{K} \rVert^2}{\langle \hat{\mathbf{n}}^{K}, (\mathbf{N} - \mathbf{K})\rangle}$ (see Fig.~\ref{fig:skeleton-shapes}, left).  The final amplitude is $a'-\epsilon$, to preserve a free space between the curves. This procedure is applied recursively, as several intersections may occur.
	
	We perform this intersection test for the translated knot points, \ie the sin peaks or the sawtooth and rectangle corners, and additionally for the midpoints of the rectangular shape, which represent the most protruding parts of the modified curves.  While our strategy successfully prevents intersections in almost all cases we observed, extreme cases, \eg, highly curved icons, which are less suited for GfIs, may require manually increasing the skeleton pixel offset $\epsilon$.
	\paragraph*{Shape} is controlled by means of additional knots and by shifting the translated positions along the curve (see Fig.~\ref{fig:skeleton-shapes}, right).
	In the case of a sinusoidal wave, we place four knots per period: two `peak knot points' (orange circles and green diamonds in Fig.~\myref{fig:params}{:5a}) and two intermediate knot points  (blue crosses in Fig.~\myref{fig:params}{:5a}), which do not function as curve modification constraints. This procedure restricts the influence of the control points to one period and guarantees a smooth shape.
	For sawtooth-like shape of a B-Spline $k(u)$, each next but one knot point $k(t_i)$ is translated along the corresponding curve normal, while the same shift in the opposite direction yields the new position of the respective next neighbor knot point $k(t_{i+1})$ (see Fig.~\ref{fig:skeleton-shapes}, right). 
	For rectangular shape, for each existent knot $t_i$ and its respective curve position $k(t_i)$, we insert a new knot $t'_i$ that corresponds to the same curve position plus a small offset. Then these pairs of knot points are translated in the opposite directions, \ie along the normal and the negative normal vector (see Fig.~\ref{fig:skeleton-shapes}, right) 
	For the sawtooth-like and rectangular shape, the knot multiplicity is increased to 3 to achieve the $C^0$-continuity and thus to produce sharp corners. After a curve modification, each third control point, starting with the first one, is located in a corner, while we place the inner two control points on the lines between the corners to get straight line strips.

	\subsubsection{Color}
	\label{s:pipeline.var.color}
	We apply color modifications to the icon's \emph{contour region} in a similar fashion as for the geometry. We create a periodic color modification along the icon contours by alternating equidistant intervals of the original and a modified luminance. While, in general, various color alteration methods are possible, we select an approach that produces a broken line like pattern, as moderate color variations are hard to distinguish. The contour region is defined by a margin width. We also modify the color of the complementary region, \ie the \emph{inner region} of the icon by applying a user-defined color map.

	\paragraph*{Color frequency (contour region)} is controlled by arc length of the color intervals. It is determined by the positions of the color attributes that serve as start and end color of the respective interval. To construct the corresponding color points, we initially compute virtual interval boundaries as equidistant arc length positions along the curve, according to the target frequency. This is done analogously to the knot vector calculation for arc-length parametrization (see Sec.~\ref{s:pipeline.var.geometry}). Afterwards, we place two new color point locations for each interval boundary with a small arc length offset to the left and right side along the curve to get a hard transition between the original and the modified color. (see Fig.~\myref{fig:params}{:4b}). The color of each second color point pair is modified to the color amplitude (see below), while the other pair stays unchanged (see Fig.~\myref{fig:params}{:5b}).
	
	\paragraph*{Color amplitude (contour region)} is controlled by increasing or decreasing the luminance of the original icon color, depending on whether the original luminance is below or above 50\%, respectively. Increasing the amplitude will gradually increase or decrease the icon color luminance until a maximum or minimum luminance is reached.
	For example, in the bi-chrome icon in Fig.~\myref{fig:params}{:5b}, the light gray circles represent the modified luminance values, while the outer color is masked out and remains white.
	
	\paragraph*{The margin width} defines the segmentation of the icon's contour and inner regions. We add diffusion barriers to the GfI representation inside the icon with the offset of the given margin width from the original contour curves. The diffusion barriers are defined as isolines of the distance field inside the icon for the given distance. 
	We extended the DCI renderer of Jeschke~\cite{jeschke2009gpu} by adding a \emph{barrier flag}, which blocks the contribution of the corresponding DC to the diffusion process. Generally, we insert one-side barriers, which limit the margin color modification to the contour region.
	
	\paragraph*{Inner Color} is modified using a perceptually uniform sequential colormap, where we commonly select a colormap that relates the icons original color. The color is applied uniformly to the inner region of the icon without any further modification using, \eg, textures.
	
	\subsection{Post-Processing}
	\label{s:pipeline.postproc}
	After a GfI has been constructed by generating the corresponding visual variables, the glyph's representation is back-converted into a DCI. The modified B-splines' de~Boor points are transformed to B\'{e}zier control points using a basis transformation matrix~\cite{casciola2004genconversmatrix}. Accordingly, the parametric positions of the color attributes are remapped onto the intervals of the resulting B\'{e}zier splines.  For rendering the DCIs as raster images, we use an extended rendering tool of Jeschke~\cite{jeschke2009gpu} as described above.

	\section{Perceptual Model}
	\label{s:model}
	To allow a quantitative visualization with GfIs, we evaluate the following perceptual aspects using an online survey:
	\begin{compactdesc}
		\item[Quantization] of visual variables, \ie definition of clearly distinguishable and perceptually equidistant magnitude levels.
		\item[Stimulus-to-perception transformation function,] \ie transformation between stimulus magnitudes and perceived values, which is often non-linear~\cite{Stevens1957PsychophysLaw}.
		\item[Shape-dependency] for geometric amplitude and frequency. It is to assume that the shape of a geometric modification influences the perception of the geometric amplitude and frequency. Thus, taken the sinusoidal shape as reference, the stimulus magnitudes for other shapes that produce the same sensation need to be acquired.
	\end{compactdesc}
	
	Moreover, we propose a transfer approach to assign the magnitude quantization levels to glyphs with different sizes.

	\subsection{Design of the Experiment} 
	\label{s:model.design}
	\begin{table}[b]
		\centering
		\begin{tabular}{|c|c|c|c|c|c|} \hline
			\bfseries Visual variable & {\bfseries min}\,[mm] & {\bfseries max}\,[mm] & {\bfseries step}\,[mm]\\\hline
			Geom. amplitude      & 0.1 \{1\}  & 1.2 \{12\} & 0.1 \\\hline
			Geom. period length  & 0.8 \{12\} & 5.1 \{1\}  & 0.4 \\\hline
			Color. period length & 5.9 \{5\}  & 12.1 \{1\}  & 1.6 \\\hline
			& {\bfseries min}\,[lum.]  & {\bfseries max}\,[lum.] & {\bfseries step}\,[lum.]\\\hline
			Color amplitude      & 0.425 \{1\} & 0.85 \{5\}  & 0.10625 \\\hline
		\end{tabular}
		\caption{Metric values for the GfI experiment with the glyph's size $50\times50$~mm. The numbers in \{\}-brackets are the corresponding visual variable values (adu) used for communication in the experiment. Note, that the frequency is proportional to the inverse period length. The color amplitude values are given as luminance values in $[0,1]$.}
		\label{tab:metric}
	\end{table}

	For estimation of a perceptual model, we conducted an online survey. The survey is based on glyphs generated with our GfI concept and comprises two main categories of experiments: \emph{magnitude estimation} and \emph{shape-dependent calibration}.
	\begin{description}[leftmargin=0cm]
		\item[Magnitude estimation.]  For each visual variable to be estimated, the available magnitude range was presented to participants as the minimum and the maximum reference GfI, captioned with the corresponding stimulus values, and a test GfI, for which they had to assign the perceived magnitude.
		
		\item[Shape-dependent calibration] To reduce the number of questions, all magnitude estimation experiments for geometric visual variables are done with the sinusoidal shape, and the ''calibration'' function that maps the stimulus amplitude and frequency of a rectangular and sawtooth-like shape against the respective parameters of a sinusoidal shape was  assessed separately. In this additional experiment, the participants had to select one out of five glyphs with the perceptually most similar magnitude to a presented sinusoidal reference. 
	\end{description}
	
	73 participants conducted the online survey and each participant provided 90 assessments of the relationship between stimulus and perception, giving a total of 6,570 data points. Tab.~\ref{tab:metric} gives the metric values for our experiments with the glyph's size $50\times50$~mm.

	\subsection{Survey Evaluation and Results}
	\label{s:model.survey}
	
\begin{figure}[bt]
		\centering
		\subfloat[]{
			\includegraphics[width=0.24\textwidth, height=0.2\textheight]{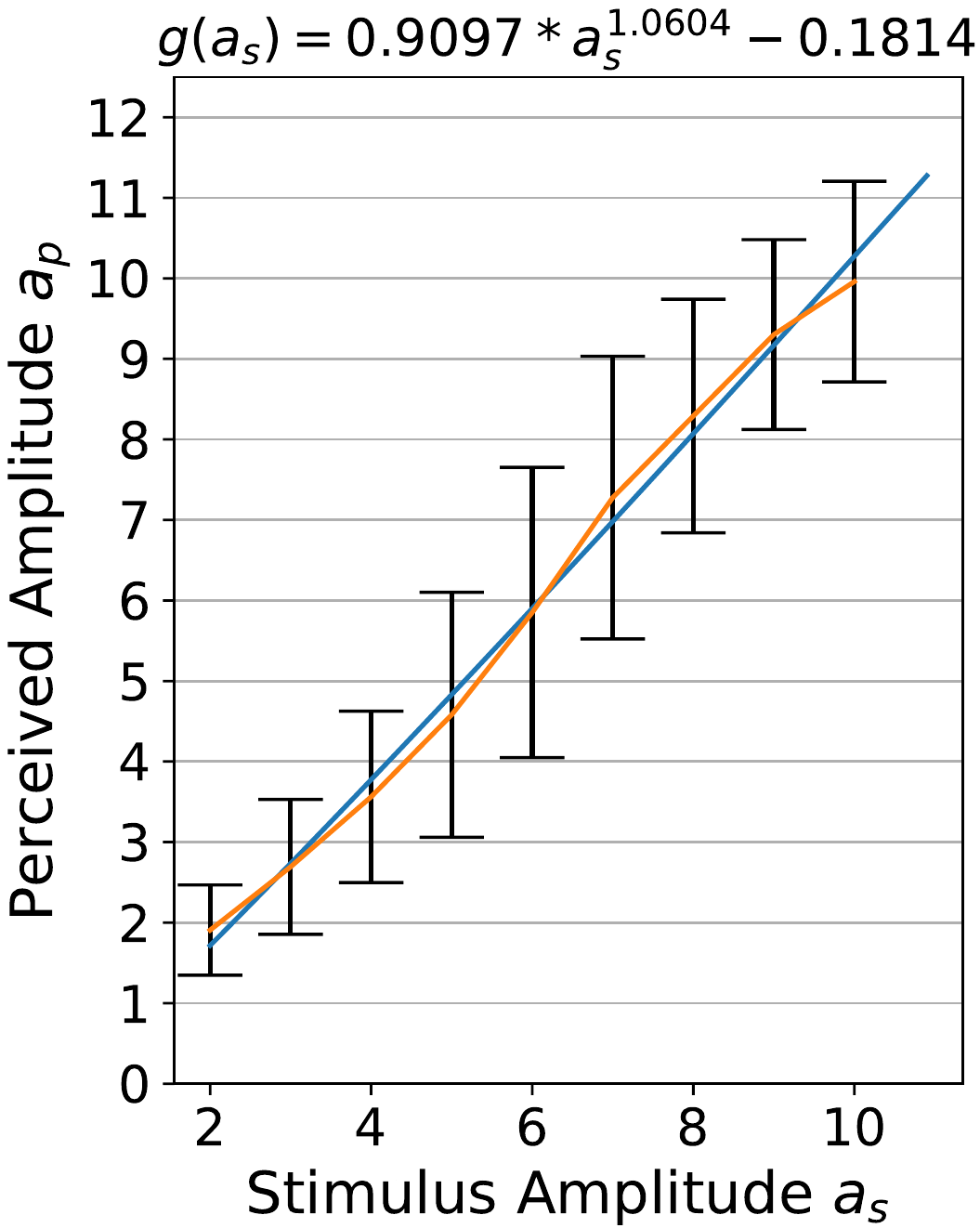} 
			\label{fig:amplFitting}
		}
		\subfloat[]{ \includegraphics[width=0.24\textwidth, height=0.2\textheight]{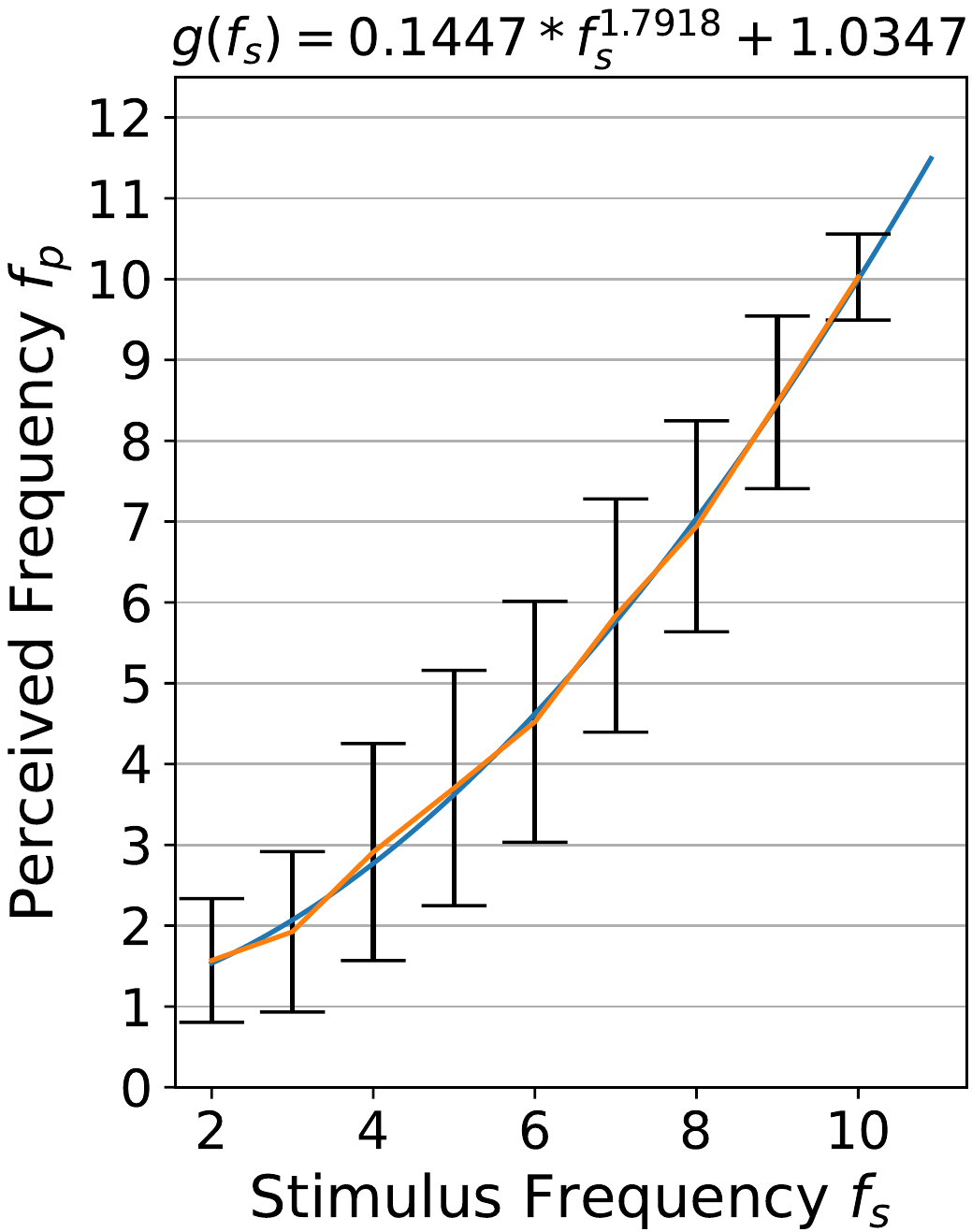}
			\label{fig:freqFitting}
		}
		\subfloat[]{
			\includegraphics[width=0.24\textwidth, height=0.2\textheight]{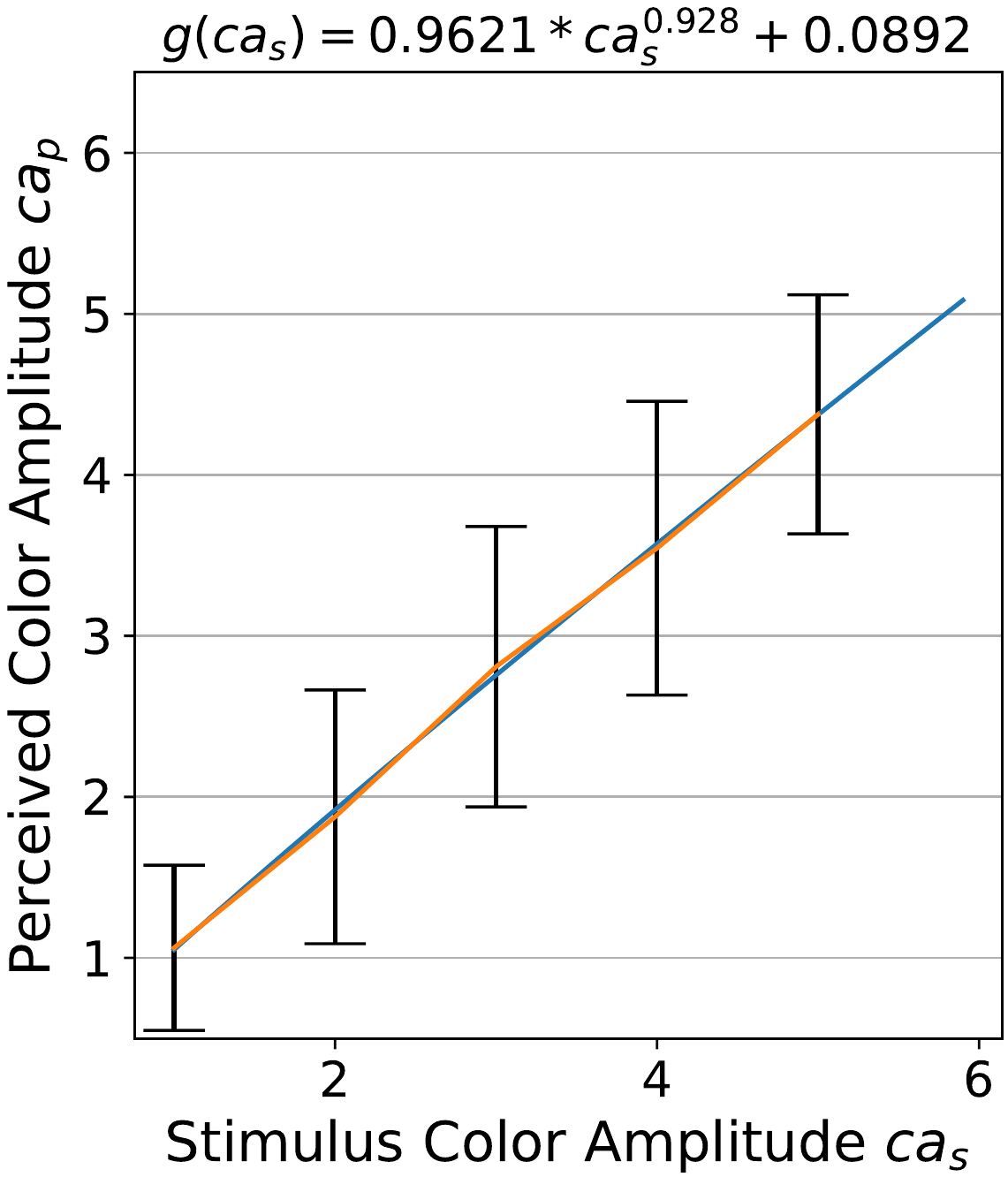} 
			\label{fig:colAmplFitting}
		}
		\subfloat[]{
			\includegraphics[width=0.24\textwidth, height=0.2\textheight]{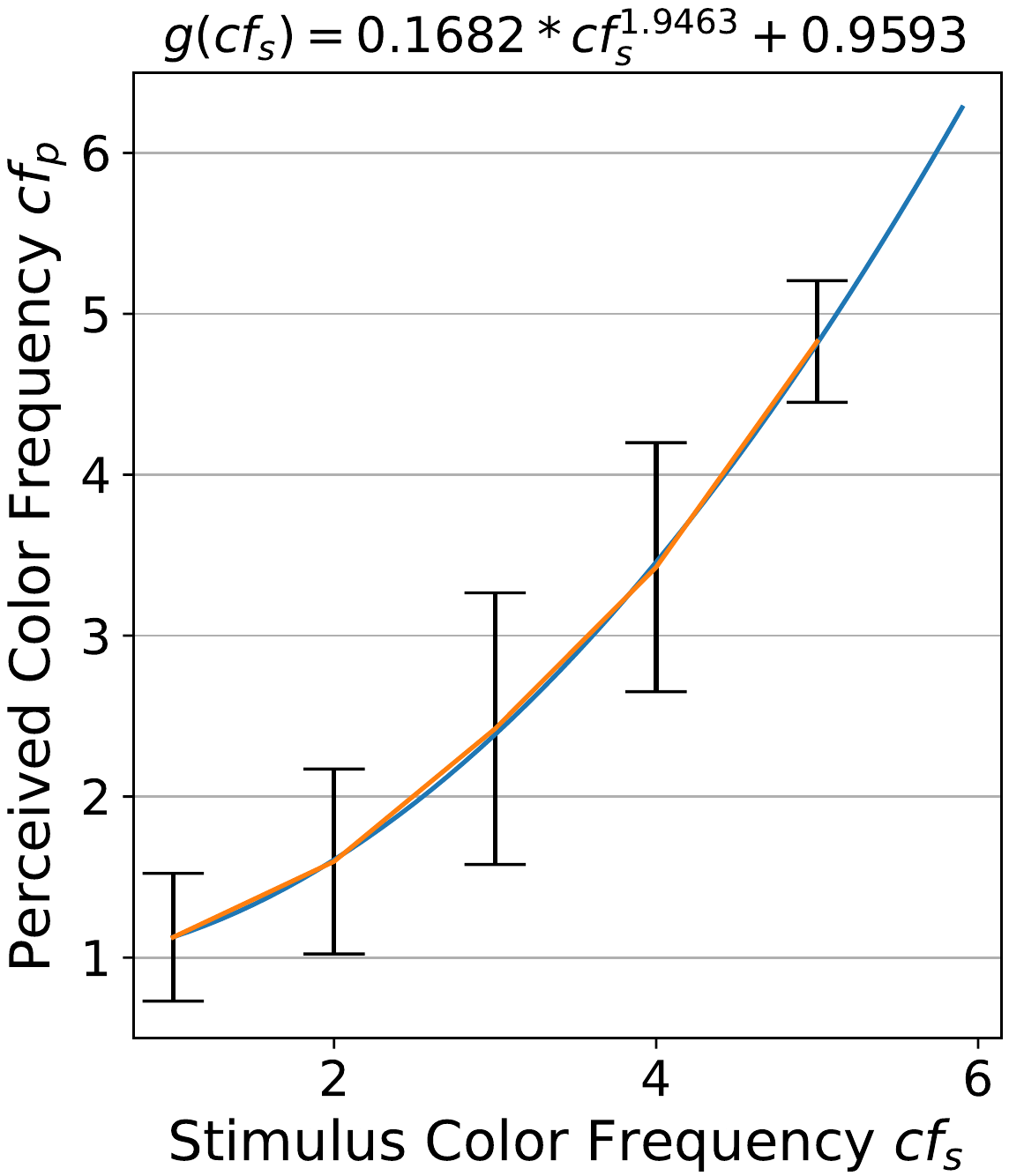} 
			\label{fig:colFreqFitting}		
		}\newline
		\subfloat[]{
			\includegraphics[width=0.24\textwidth, height=0.2\textheight]{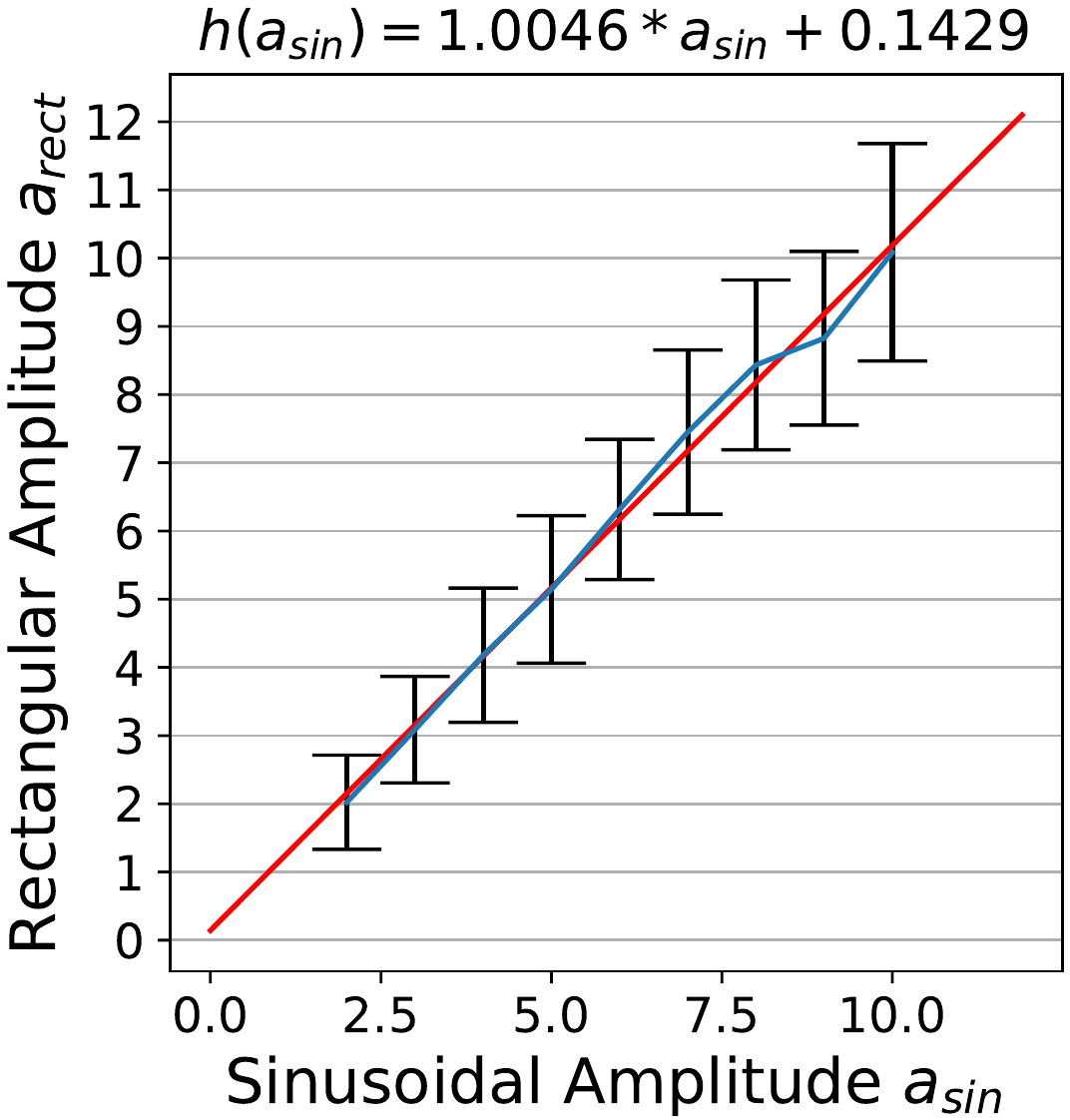} 
			\label{fig:amplCalibrRect}
		}
		\subfloat[]{
			\includegraphics[width=0.24\textwidth, height=0.2\textheight]{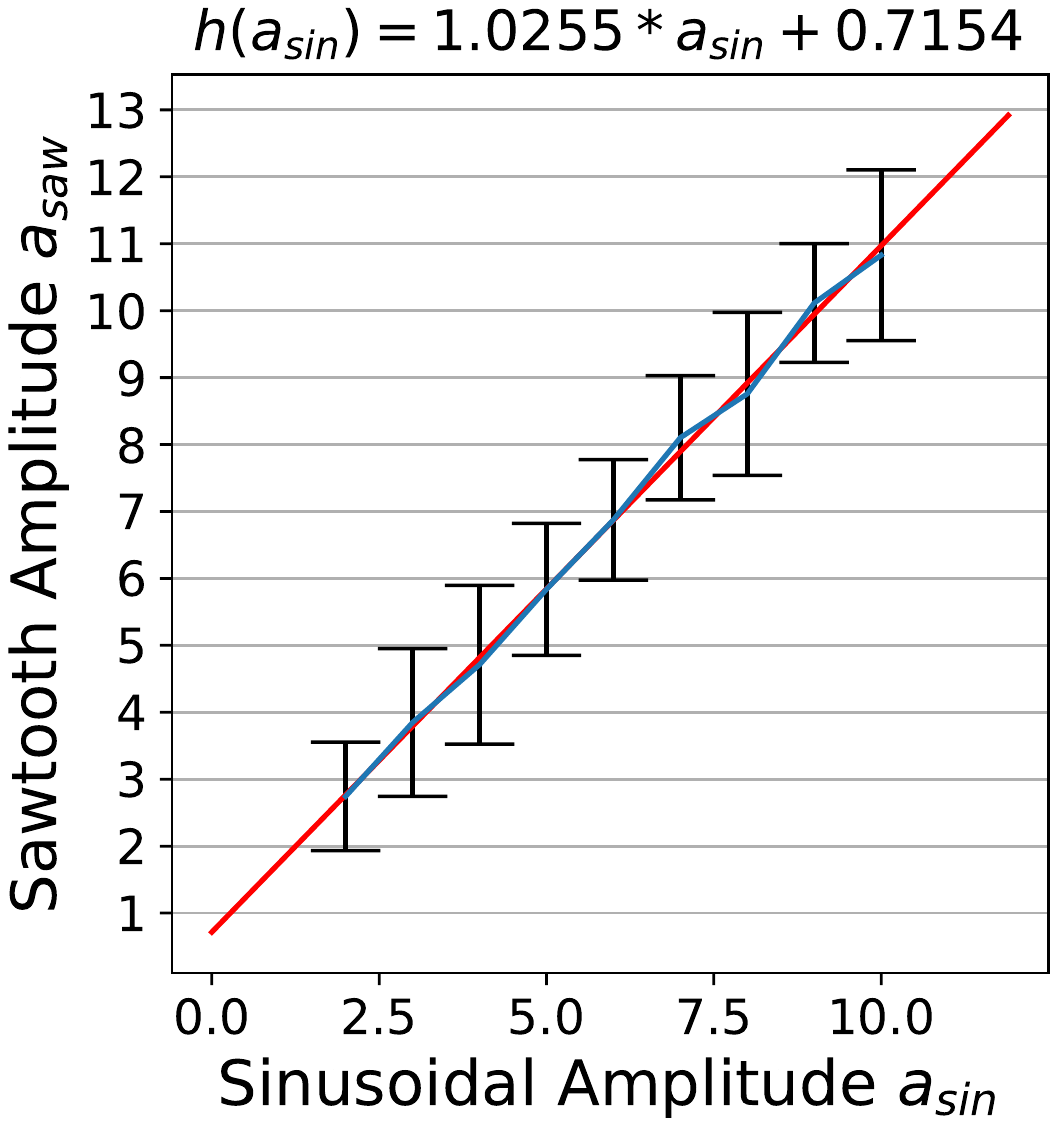} 
			\label{fig:amplCalibrSaw}
		}
		\subfloat[]{
			\includegraphics[width=0.24\textwidth, height=0.2\textheight]{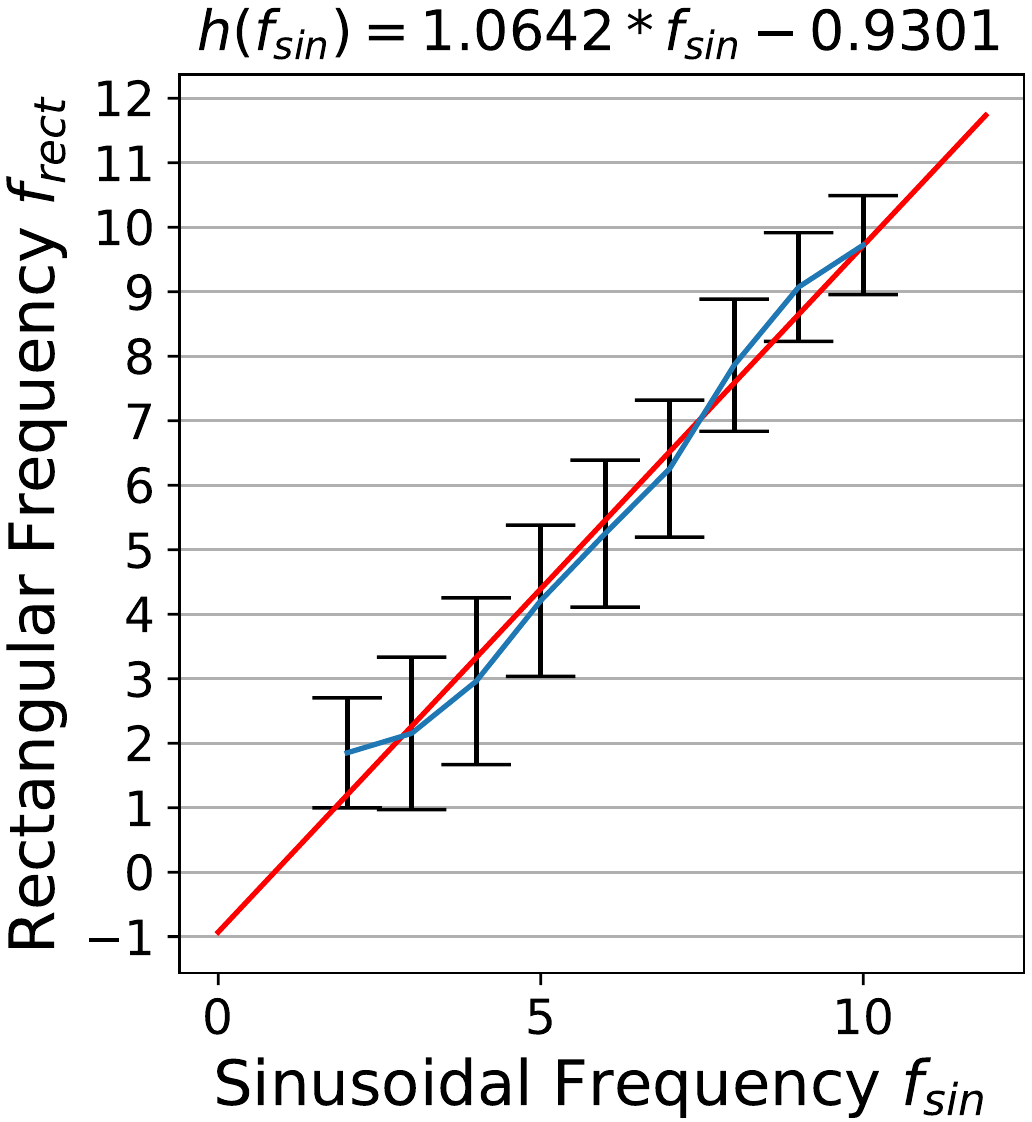} 
			\label{fig:freqCalibrRect}
		}
		\subfloat[]{
			\includegraphics[width=0.24\textwidth, height=0.2\textheight]{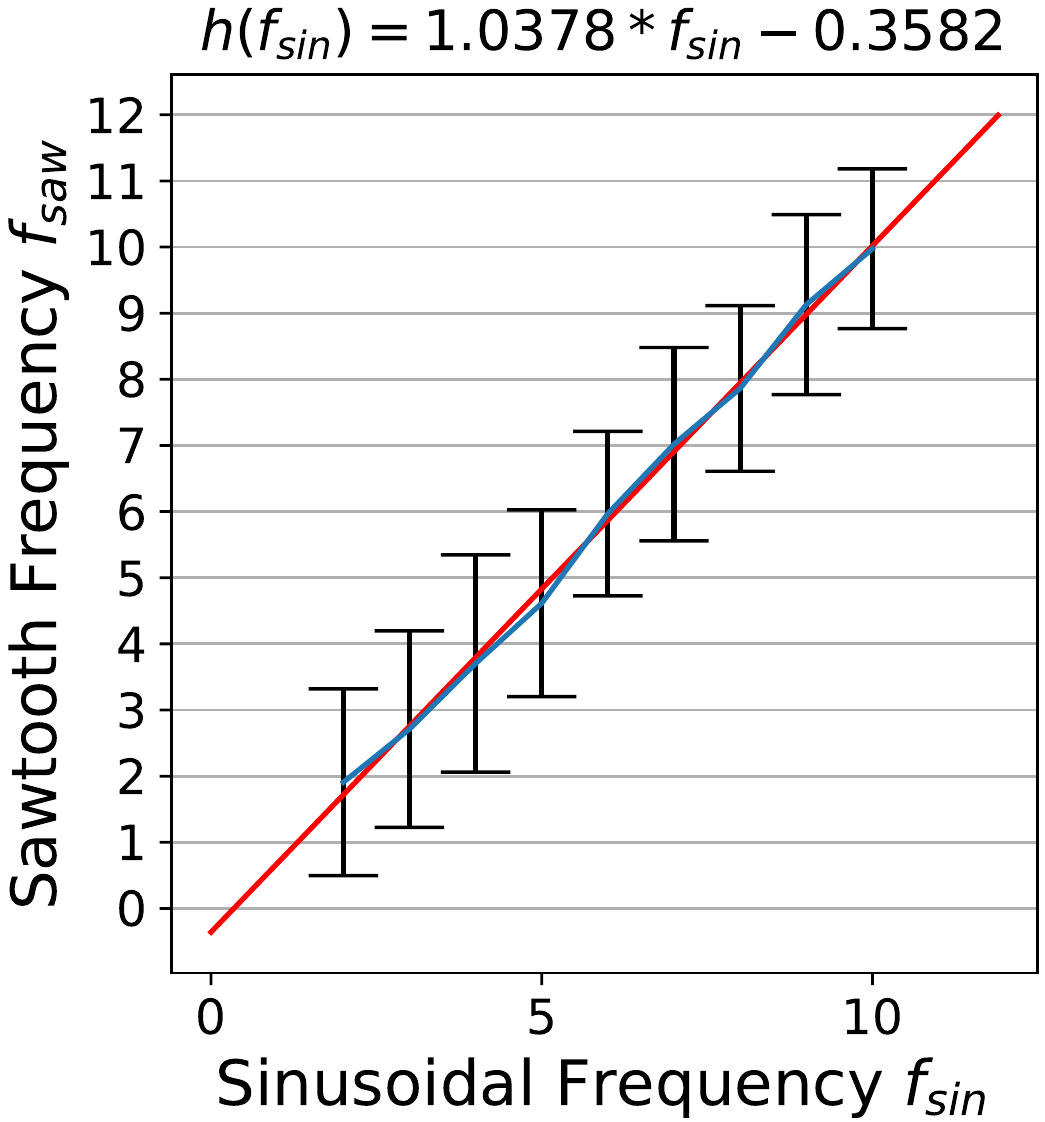} 
			\label{fig:freqCalibrSaw}		
		}
		\caption{Top: Modeling stimulus-to-perception transformation for geometric amplitude (\ref{fig:amplFitting}), geometric frequency (\ref{fig:freqFitting}), color amplitude (\ref{fig:colAmplFitting}), and color frequency (\ref{fig:colFreqFitting}) including the fitted transformation functions (blue) and perceptual means (orange).\newline Bottom: Calibration of rectangular and sawtooth-like shape against sinusoidal shape for geometric amplitude of rectangular (\ref{fig:amplCalibrRect}) and sawtooth-like shape (\ref{fig:amplCalibrSaw}), and frequency of rectangular (\ref{fig:freqCalibrRect}) and sawtooth-like shape (\ref{fig:freqCalibrSaw}).}
		\label{fig:transfFitting_shapeCalibr}
	\end{figure}

	\paragraph*{Modeling the stimulus-to-perception transformation function.}	
	Following Stevens~\cite{Stevens1957PsychophysLaw}, we assume that the stimulus-to-perception transformation has the form of a power function $g(\mathbf{x})=a\cdot\mathbf{x}^b + c$. Thus, having the perceived magnitudes as data points $g(\mathbf{x})$, as stated by the participants, and the stimulus magnitudes $\mathbf{x}$ as the independent parameter, $a$, $b$ and $c$ are estimated using non-linear least-square fitting. The modeled functions are presented in Fig.~\ref{fig:transfFitting_shapeCalibr}, top. For both, geometry and color, we observe a rather linear and positive power dependency of the perceived magnitudes from the stimulus magnitudes for amplitude and frequency, respectively.
	
	\paragraph*{Quantization.}
	The aim is to find a quantization step such that all resulting magnitude levels do not overlap with neighboring confidence intervals for a given confidence level. This is analogous to the principle applied by estimation of just noticeable difference (jnd), which is defined regarding the probability of correct assignments, usually 50\%, which we also apply in our experiment. Tab.~\ref{tab:quant} gives an overview of the quantization steps and the resulting number of discrete levels for each visual variable. Further statistics of the quantization process and details of the computation method are given in the supplementary material.
	
	\paragraph*{Shape-dependent calibration.}
	The shape calibration is modeled as a linear function $h(\mathbf{x})=k\cdot\mathbf{x}+l$, with $\mathbf{x}$ being the reference sinusoidal magnitudes from the calibration experiments (see Sec.~\ref{s:model.design}) and the perceived magnitudes regarding the rectangular or sawtooth-like shape as data points. The fitting of $h$ is done with a least-square method. Fig.~\ref{fig:transfFitting_shapeCalibr}, bottom, shows the resulting functions per shape and stimulus. These results demonstrate that the influence of a specific shape on the perception of amplitude and frequency magnitudes is rather marginal.

	\begin{table}[bt]
		\centering
		\begin{tabular}{|c|c|c|c|c|}\hline
				\textbf{Vis. Variable} & \multicolumn{2}{c|}{\textbf{Geometry}} & \multicolumn{2}{c|}{\textbf{Color}}\\
				& Ampl. & Freq .& Ampl. & Freq.\\\hline
				Quant. step $\Delta v$ (adu)& $\approx 2.91$ & $\approx 2.01$ & $\approx 1.23$ & $\approx 1.14$  \\\hline
				\# levels ($50$~mm)& 4 & 5 & 4 & 4  \\\hline	
				\# levels ($16$~mm)& 3 & 3 & 4 & 3  \\\hline
				$\omega_{\text{min}}/\omega_{\text{max}}$ Fig.\ref{fig:incHosp},\ref{fig:inchospDeath},\ref{fig:bedsVent} & 0.558/0.66 & 0.558/0.43 & n.a. & 0.558/0.43  \\\hline
				$\omega_{\text{min}}/\omega_{\text{max}}$ Fig.\ref{fig:beds} & 0.728/0.69 & 0.728/0.45 & n.a. & 0.728/0.45  \\\hline
		\end{tabular}
		\caption{Quantization results and scaling parameters: Quantization step-sizes and number of levels derived from the user experiment with the glyph's size $50\times50$~mm, the number of levels after transferring to the glyph size $16\times16$~mm, and corresponding scaling parameters.}
		\label{tab:quant}
	\end{table}

	\subsection{Transfer to Different Glyph Sizes}
	\label{s:model.scale}
	From the survey and the evaluation in Sec.~\ref{s:model.survey}, we derived the levels of visual variables for the geometry and the color for a fixed glyph size $50\times50$~mm. More precisely, for a perceived visual variable $v$ we initially fixed the minimum and the maximum stimulus values $s_{\text{min}}, s_{\text{max}}$ (in mm) and deduced the quantization size $\Delta v$ (in arbitrary digital units, adu) applied to the range $v_{\text{min}}, v_{\text{max}}$ (in adu) from the user experiment, which corresponds to $s_{\text{min}}$ and $s_{\text{max}}$ (see Sec.~\ref{s:model.survey}). Note, that $\Delta v$ corresponds to $\Delta s(v)$ (in mm), which is in general not constant (see Fig.~\ref{fig:transfFitting_shapeCalibr}).
	
	To transfer this information to an icon of different size with a relative scale $\omega>0$ to the original glyph of $50\times50$~mm, we apply the following rules (we assume $0<\omega<1$ in the following, since glyphs are commonly used at smaller scales)
	\begin{compactitem}
		\item Color amplitude should not be scaled, as luminance is independent of size.
		\item The ``perceptual'' stepsize $\Delta v$ (and the corresponding stimulus stepsizes $\Delta s(v)$) should not be reduced to preserve the absolute variation (in mm) and thus the visual distinguishability.
		\item The minimum and the maximum stimulus and visual variable values $s_{\text{min}}, s_{\text{max}}$ and $v_{\text{min}}, v_{\text{max}}$, respectively, are scaled according to the following rules
		\begin{compactitem}
			\item The minimum values $s_{\text{min}}$ and $v_{\text{min}}$ can only be scaled moderately, \ie reduced using $\omega_{\text{min}}>\omega$, maybe even $\omega_{\text{min}}=1$, to prevent, for example, visually vanishing amplitudes.
			\item The maximum values $s_{\text{max}}$ and $v_{\text{max}}$ should be scaled by $\omega_{\text{max}}=\omega$, to prevent, for example, extreme distortions for small icons.
		\end{compactitem}
		Consequently, the number of levels gets potentially reduced for $\omega<1$ as the ``usable'' range $[\omega_{\text{min}}\cdot s_{\text{min}},\omega_{\text{max}}\cdot s_{\text{max}}]$ gets smaller while the stepsize $\Delta s(v)$ remains unchanged. To counteract on this problem, we reduce the scaling effect for the maximum values $\omega_{\text{max}}=\omega+\epsilon\leq1$ with a user-defined parameter $\epsilon$ that also depends on the icon's geometric complexity. 
	\end{compactitem}

	\section{Application Examples}
	\label{s:appl}
	The starting point for a practical usage of our GfIs is the selection of a representative icon, the proper scale (see Sec.~\ref{s:model.scale}), and the mapping of the application data to the up to seven glyph variables.
	While our user study, presented in Sec.~\ref{s:model.design} and \ref{s:model.survey}, evaluated one or two variables at once, the parallel use of the more visual variables further reduces the distinguishability and readability of the individual variable values. We, therefore, propose to combine variables to enhance the readability in multidimensional settings.
	
\begin{figure*}[tb]
		\centering
		\subfloat[``2D'' virus glyph]{
			\includegraphics[width=0.49\textwidth]{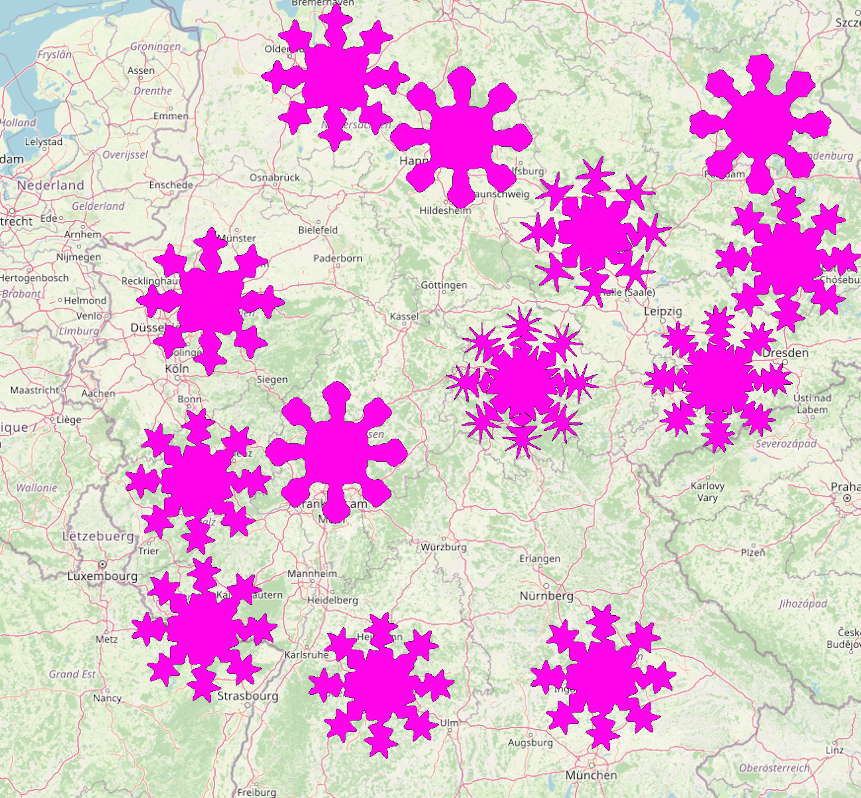} 
			\label{fig:incHosp}
		}
		\subfloat[``3D'' virus glyph]{
			\includegraphics[width=0.49\textwidth]{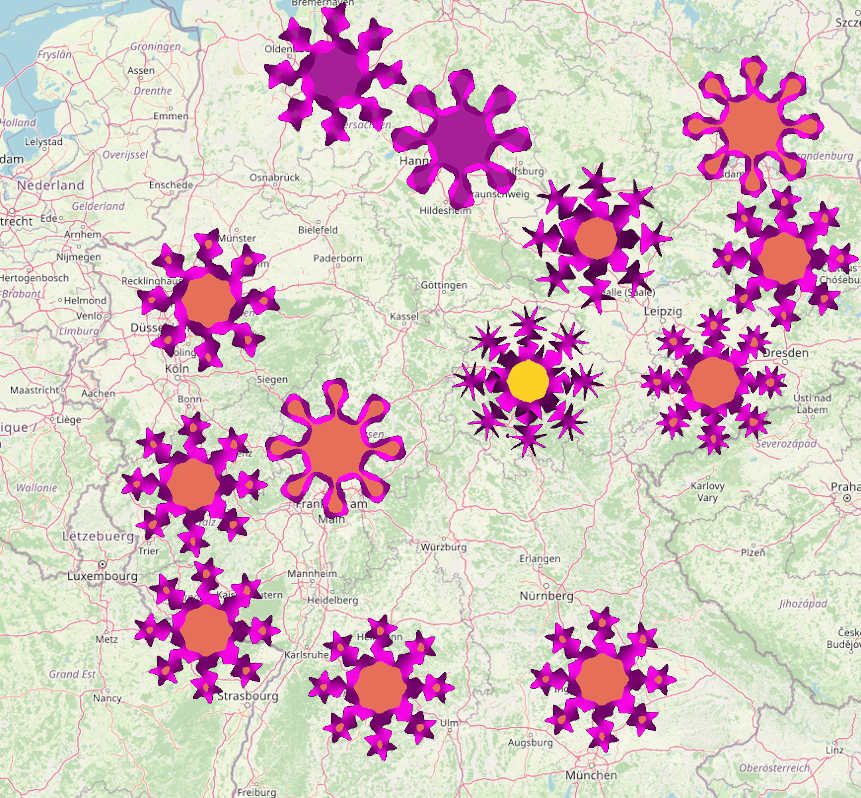} 
			\label{fig:inchospDeath}
		}\\
		\subfloat[``2D'' hospital-shaped glyph]{
			\includegraphics[width=0.49\textwidth]{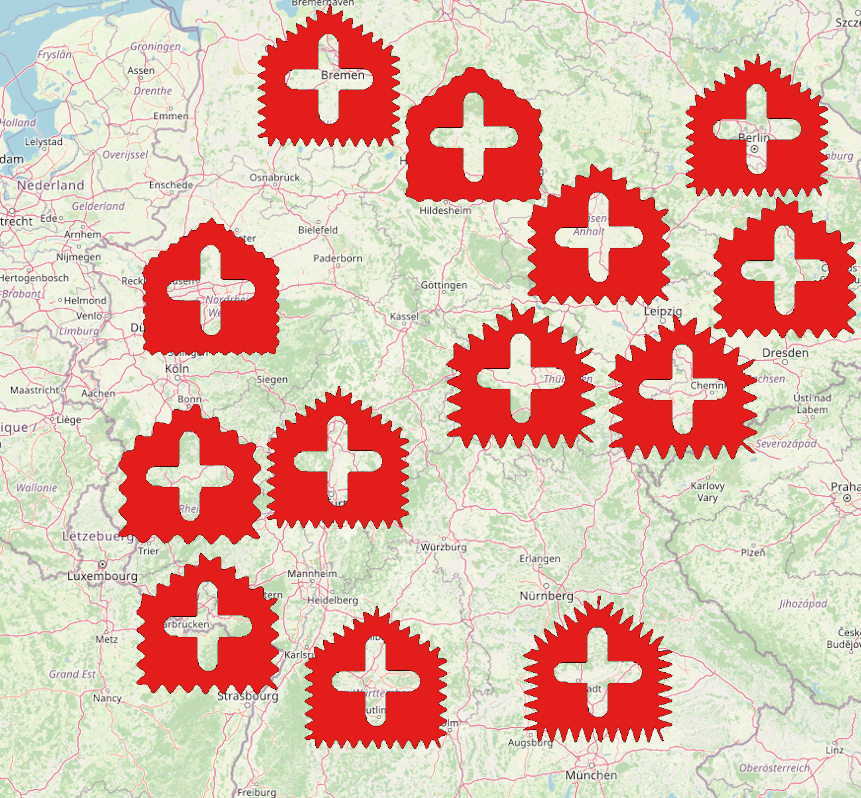} 
			\label{fig:beds}
		}
		\subfloat[``2D'' lung glyph]{
			\includegraphics[width=0.49\textwidth]{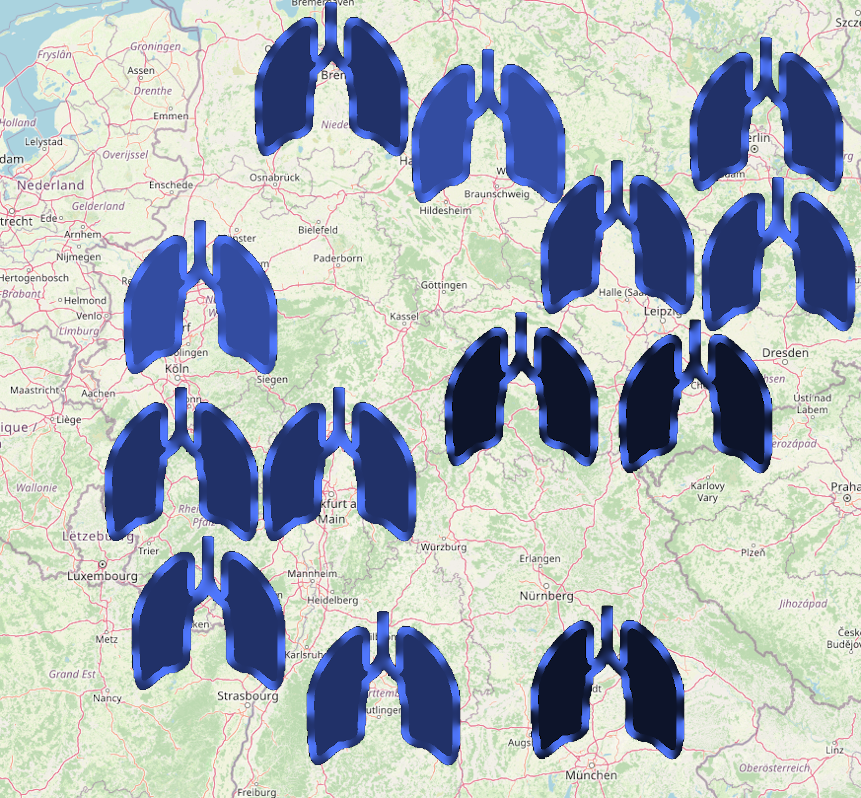} 
			\label{fig:bedsVent}
		}
		\caption{COVID-19 situation in central and southern Germany. Fig.~\ref{fig:incHosp} encodes last 7 days cases / 100K pop. as frequency, last 7 days hospitalizations / 100K pop. as amplitude. Fig.~\ref{fig:inchospDeath} encoding: last 7 days cases / 100K pop. as frequency, last 7 days hospitalizations /1 00K pop. as amplitude, deaths / 100K pop. as color amplitude. Fig.~\ref{fig:beds} encoding: free intensive care beds / total intensive care beds as frequency, COVID-19 patients / intensive care beds as amplitude. Fig.~\ref{fig:bedsVent} encoding: COVID-19 patients / intensive care beds as color amplitude, COVID-19 patients ventilated / COVID-19 patients total as color frequency. Map courtesy by OpenStreetMap, Germany.}
		\label{fig:covid}
	\end{figure*}

	In the following we give four examples to visualize COVID-19 related data from the end of Nov. 2021 for the central and southern part of Germany; see Fig.~\ref{fig:covid}. Two of the examples use visual variables directly and two use combined variables. All presented glyphs are generated for a size of $16\times16$~mm. The scaling factors  are presented in Tab.~\ref{tab:quant}.
	The corresponding mapping of the COVID-19 data to the visual variables and the visual legends are given Tab.~\ref{tab:encod}.

	Figs.~\ref{fig:incHosp} and \ref{fig:inchospDeath} visualize a data set from a COVID-19 Daily Situation Report by the RKI~\cite{RKI-cases} for Nov.~30,~2021. Fig.~\ref{fig:incHosp} depicts the 7-days cases incidence value and the 7-days hospitalization value as frequency and amplitude of a sinusoidal contour shape, respectively, using a virus-shaped icon as base geometry, which is also used in Fig.~\ref{fig:concept}. Fig.~\ref{fig:inchospDeath} uses the same icon, but applies combined visual variables, \ie geometric and color frequency jointly represent 7-days cases incidence value, while geometric and color amplitude and the width of the contour region encode the 7-days hospitalization. Additionally, the inner color represents the deaths per 100K population.
	
	Fig.~\ref{fig:beds} visualizes the intensive care bed occupancy in Germany at Nov.~30,~2021~\cite{RKI-beds} using a hospital-shaped icon. Here, the relative number of free intensive care beds and the relative number of COVID-19 patients per intensive care beds are encoded as sinusoidal frequency and amplitude, respectively. The example in Fig.~\ref{fig:bedsVent} visualizes the relative number of COVID-19 patients per intensive care beds as combined visual variable using luminance amplitude and and inner color, and the percentage of ventilated COVID-19 patients relative to the total number of COVID-19 patients as color frequency, using a lunge icon as the base geometry.
	
	Further examples including colored icons are given in the supplementary material.

	\begin{table}[tb]
		\renewcommand{\tabcolsep}{4pt}
		\centering
		\begin{tabular}{|c|>{\centering\arraybackslash}p{.23\textwidth}|>{\centering\arraybackslash}p{.15\textwidth}|>{\centering\arraybackslash}p{.12\textwidth}>{\centering\arraybackslash}p{.13\textwidth}>{\centering\arraybackslash}p{.12\textwidth}|}\hline
			\textbf{Fig} & \textbf{Data} & \textbf{Vis. Var.} & \multicolumn{3}{c|}{\textbf{Mapping}}\\\hline
			\ref{fig:incHosp}& 7 day cases / 100K popul. & Geometric frequency &
			$\left[0,300\right[$    \includegraphics[width=5.5mm]{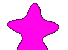} &
			$\left[300, 900\right[$ \includegraphics[width=5.5mm]{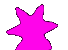} &
			$\geq 900$              \includegraphics[width=5.5mm]{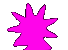} 
			\\\hline
			\ref{fig:incHosp}&7 day hospital. / 100K popul. & Geometric amplitude &
			$\left[0,4\right[$   \includegraphics[width=5.5mm]{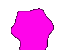} &
			$\left[4,10\right[$  \includegraphics[width=5.5mm]{virus_0-65_sinF3A2_legend.png} &
			$\geq 10$            \includegraphics[width=5.5mm]{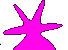} 
			\\\hline
			\ref{fig:inchospDeath}& 7 day cases / 100K popul. & Comb. frequency &
			$\left[0,300\right[$    \includegraphics[width=5.5mm]{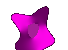} &
			$\left[300, 900\right[$ \includegraphics[width=5.5mm]{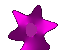} &
			$\geq 900$              \includegraphics[width=5.5mm]{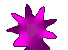} 
			\\\hline
			\ref{fig:inchospDeath}&7 day hospital. / 100K popul. &
			Comb. amplitude &
			$\left[0,4\right[$   \includegraphics[width=5.5mm]{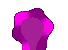} &
			$\left[4,10\right[$  \includegraphics[width=5.5mm]{virus_c1_sinF2A2_legend.png} &
			$\geq 10$            \includegraphics[width=5.5mm]{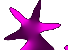} 
			\\\hline
			\ref{fig:inchospDeath}&Deaths / 100K popul. & Inner color &
			$\left[0,100\right[$    \includegraphics[width=5.5mm]{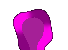} &  
			$\left[100, 200\right[$ \includegraphics[width=5.5mm]{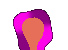} &  
			$\geq 200$              \includegraphics[width=5.5mm]{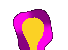}    
			\\\hline
			\ref{fig:beds}&Free intensity care beds [\%] & Geometric frequency &
			$>15$                   \includegraphics[width=5.5mm]{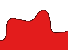} &    
			$\left]10, 15\right]$   \includegraphics[width=5.5mm]{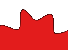} &    
			$\left[0, 10\right]$    \includegraphics[width=5.5mm]{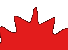} 
			\\\hline
			\ref{fig:beds}&COVID-pat. / intens. beds & Geometric amplitude &
			$\left[0,15\right[$    \includegraphics[width=5.5mm]{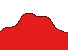} &
			$\left[15, 30\right[$  \includegraphics[width=5.5mm]{hospital_sinF2A2_legend.png} &
			$\geq 30$              \includegraphics[width=5.5mm]{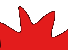}
			\\\hline
			\ref{fig:bedsVent}&COVID-pat. / intens. beds & Comb. amplitude&
			$\left[0,15\right[$   \includegraphics[width=5.5mm]{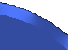} &      
			$\left[15, 30\right[$ \includegraphics[width=5.5mm]{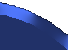} &      
			$\geq 30$             \includegraphics[width=5.5mm]{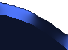} 
			\\\hline
			\ref{fig:bedsVent}& Ventilation COVID-pat [\%] & Color frequency &
			$\left[0,45\right[$   \includegraphics[width=5.5mm]{lung_colCA3CF1_legend.png} &    
			$\left[45, 55\right[$ \includegraphics[width=5.5mm]{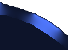} &    
			$\geq 55$             \includegraphics[width=5.5mm]{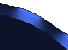}      
			\\\hline
		\end{tabular}
		\caption{Visual encoding of COVID-19 data.} 
		\label{tab:encod}
	\end{table}

	\section{Discussion \& Conclusion}
	\label{s:discuss}
	This paper introduces the Glyph-from-Icon (GfI) concept for automated generation of metaphoric glyphs from arbitrary icons by modifying seven visual variables, namely geometric shape, amplitude and frequency as well as color amplitude and frequency, contour region width and inner color of the icon. Our approach transforms icons into an extended version of diffusion-curve using B-splines, which allows us to add the required degrees-of-freedom for the  periodic, wave-like modifications.
	We performed a user study to evaluate the perception of the visual variables of the generated glyphs, and to estimate the relation between perception and stimuli values and the quantization levels that ensure perceptual monotonicity and readability of the GfI.
	Moreover, we present rules for applying the generation method to icons with different size. 
	
	\paragraph*{Limitations.} 
	As discussed in Sec.~\ref{s:model.design}, we did not evaluate the independence between the geometric and the color variables, as this would have overloaded the survey due to the combinatorial explosion.
	Conceptually, the GfI concept can not be applied to highly detailed icons or icons dominated by thin geometries or short diffusion curve strokes, \ie in cases where applying GfI modifications results in a destruction of the basic shape and color appearance of the provided icon.  
	Moreover, compared to more specialized approaches to metaphoric glyphs, such as RoseShapes~\cite{cai2015applying}, the number of readable quantitative values is limited.
	
	\paragraph*{Potentials.}
	The presented approach can be extended in various ways, \eg by applying the visual variables in a local instead of a global way to emphasize certain (semantic) regions in the icon, or by generating animated glyphs using our parameterized approach.

\section*{Acknowledgments}
The work is funded by the Deutsche Forschungsgemeinschaft (DFG, German Research Foundation) – Project-ID 262513311 – SFB 1187, sub-project A06 ``Visually integrated medical cooperation''.

\bibliographystyle{abbrv-doi}

\bibliography{references}
\end{document}


\maketitle

In this supplementary material we provide some additional details regarding the design of the experiment and the evaluation of the survey results, complementing the figures of the main paper. Furthermore, we show more examples of glyphs constructed from different icons.

\section{Online Survey and Perceptual Model}
\label{s:model}
To allow a quantitative visualization with GfIs, we evaluate the following perceptual aspects:
\begin{compactdesc}
	\item[Quantization] of visual variables, \ie definition of clearly distinguishable and perceptually equidistant magnitude levels.
	\item[Stimulus-to-perception transformation function,] \ie transformation between stimulus magnitudes and perceived values, which is often non-linear~\cite{Stevens1957PsychophysLaw}.
	\item[Shape-dependency] for geometric amplitude and frequency. It is to assume that the shape of a geometric modification influences the perception of the geometric amplitude and frequency. Thus, taken the sinusoidal shape as reference, the stimulus magnitudes for other shapes that produce the same sensation need to be acquired.
\end{compactdesc}

\subsection{Design of the Experiment} 
\label{s:model.design}

\begin{figure}[h!]
	\centering
	\includegraphics[width=0.32\textwidth]{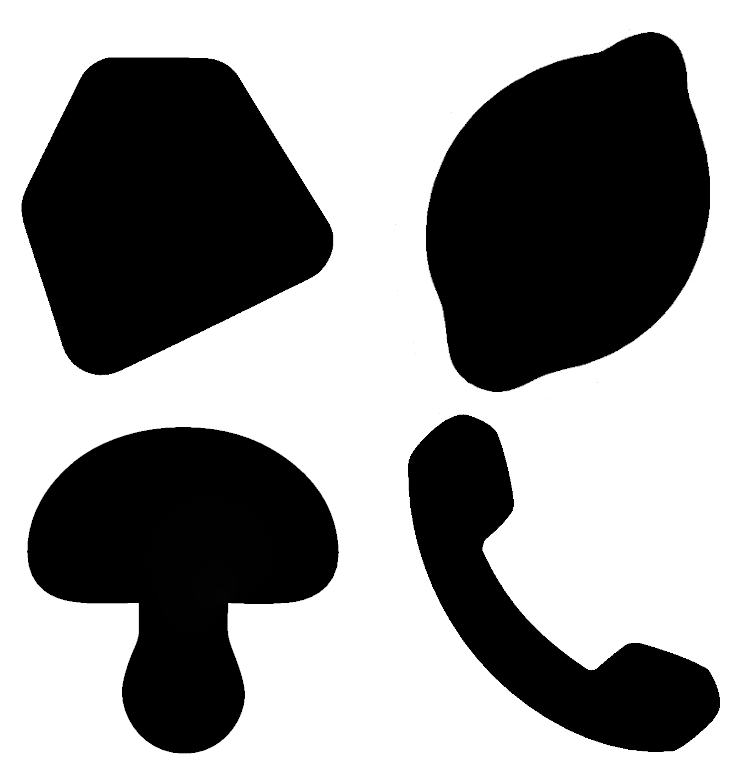} 
	
	\caption{Four base icons used for GfI generation in the experiments.}
	\label{fig:surv.icons}
\end{figure}	

\begin{figure*}[bt]
	\centering
	\begin{minipage}{.4\textwidth}
		\centering
		\subfloat[Amplitude estimation.]{
			\includegraphics[width=0.8\textwidth,valign=c]{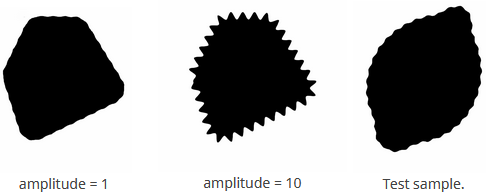} 
			\label{fig:surv.amp}
		}\\[-1mm]
		\subfloat[Frequency estimation.]{
			\includegraphics[width=0.8\textwidth,valign=c]{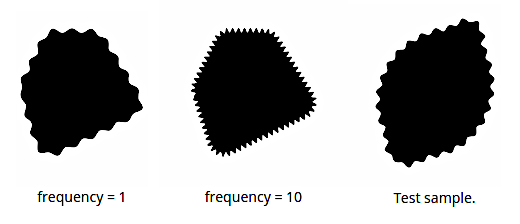} 
			\label{fig:surv.freq}
		}
	\end{minipage}\hfill
	\begin{minipage}{.55\textwidth}
		\subfloat[Amplitude estimation between shapes.]{
			\includegraphics[width=.88\textwidth,valign=c]{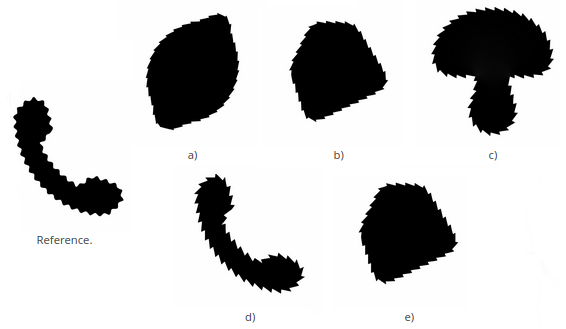}
			\label{fig:surv.saw}}
	\end{minipage}
	\caption{Examples of different geometric experiment types. Figs.~\ref{fig:surv.amp} and \ref{fig:surv.freq} are samples for direct estimation of shape amplitude (with fixed frequency) and frequency magnitude (with fixed amplitude), respectively. Fig.\ref{fig:surv.saw} is a sample for sawtooth amplitude calibration against a sinusoidal reference.}
	\label{fig:surv}
\end{figure*}

\begin{figure}[h!]
	\centering
	\subfloat[Color amplitude estimation.]{
		\includegraphics[width=0.34\textwidth]{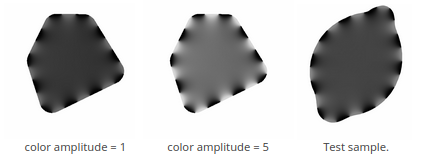} 
		\label{fig:surv.colAmp}
	} \hspace{0.22\textwidth}
	\subfloat[Color frequency estimation.]{
		\includegraphics[width=0.34\textwidth]{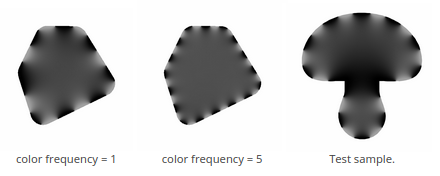} 
		\label{fig:surv.colFreq}
	}
	\caption{Examples of different color experiment types. Figs.~\ref{fig:surv.colAmp} and \ref{fig:surv.colFreq} show samples for direct estimation of color amplitude magnitude (with fixed color frequency) and color frequency magnitude (with fixed color amplitude), respectively.}
	\label{fig:surv.exp}
\end{figure}

\begin{table}[b]
	\centering
	\tiny 
	\resizebox{0.8\textwidth}{!}{
		\begin{tabular}{|c|c|c|c|c|c|} \hline
			\bfseries Visual variable & {\bfseries min}\,[mm] & {\bfseries max}\,[mm] & {\bfseries step}\,[mm]\\\hline
			Geom. amplitude      & 0.1 \{1\}  & 1.2 \{12\} & 0.1 \\\hline
			Geom. period length  & 0.8 \{12\} & 5.1 \{1\}  & 0.4 \\\hline
			Color. period length & 5.9 \{5\}  & 12.1 \{1\}  & 1.6 \\\hline
			& {\bfseries min}\,[lum.]  & {\bfseries max}\,[lum.] & {\bfseries step}\,[lum.]\\\hline
			Color amplitude      & 0.425 \{1\} & 0.85 \{5\}  & 0.10625 \\\hline
	\end{tabular}}
	\caption{Metric values for the GfI experiment with the glyph's size $50$~mm. The number in \{\}-brackets are the corresponding visual variable values (adu) used for communication in the experiment. Note, that the frequency is proportional to the inverse period length. The color amplitude values are given as luminance values in $[0,1]$.}
	\label{tab:metric}
\end{table}

The user survey is based on  glyphs generated with our GfI concept. We use four generic monochrome images with white background and black foreground (see Fig.~\ref{fig:surv.icons}). The icons as well as the generated glyphs have size $512\times512$~px and are displayed at size $50\times50$~mm. Tab.~\ref{tab:metric} summarizes the metric values used to generate the stimuli and give the mapping to the stimulus parameter values used for communication in the experiment (see also Figs.~\ref{fig:surv.amp} and \ref{fig:surv.freq}).

In each question, the base icon of the test GfIs was selected randomly. The survey comprises two main categories of experiments. 
\begin{description}[leftmargin=0cm]
	\item[Magnitude estimation.] According to the application scope of our glyph generation approach, \ie the direct reading of data values from a glyph, we performed several magnitude estimation experiments~\cite{Stevens1957PsychophysLaw} to determine a proper quantization of the visual variables as well as the transformation function between the stimuli and perception parameters. 
	
	For each visual variable to estimate, the participants got displayed the available magnitude range by presenting a minimum and maximum reference GfI with the corresponding stimulus parameter values (see Figs.~\ref{fig:surv.amp} and \ref{fig:surv.freq} for geometry and Figs.~\ref{fig:surv.colAmp} and \ref{fig:surv.colFreq} for color). The test GfIs with randomly selected magnitude were hidden by default and was uncovered for eight seconds by clicking the corresponding button, and the participants had to assign the perceived magnitude from a drop-down list. The stepsize for generating the visual stimuli for the test glyphs (see Tab.~\ref{tab:metric}) was selected to be well below the just noticeable distance to be able to derive a suitable quantization from a statistical evaluation.
	
	There are two subtypes of the magnitude estimation experiments in our survey (see Tab.~\ref{tab:questions}):
	\begin{compactenum}
		\item Fixed second stimulus, \eg shape amplitude estimation with a fixed shape frequency, and
		\item Randomly selected second stimulus.
	\end{compactenum}
	The experiments with fixed second stimulus have been placed at the beginning of the specific experiment section to make the participants acquainted with the  experimental setting, as the experiments with randomly selected second stimulus are more challenging. 
	
	\item[Shape-dependent calibration] has been performed by selecting the glyphs with the closest magnitude. To reduce the number of questions, all magnitude estimation experiments for geometric visual variables are done with the sinusoidal shape. To estimate a shape calibration function, the participants had to select one out of five glyphs with the perceptually most similar magnitude to a presented sinusoidal reference (see Fig.~\ref{fig:surv.saw}). The GfIs offered for selection were created with the magnitude levels $l \in [l^{\text{ref}}-2 \lldots l^{\text{ref}}+2]$, where $l^{\text{ref}}$ is the visual variable values (adu) we defined for the reference glyph, and have been arranged randomly. These experiments are done separately for each shape, \ie rectangular or sawtooth-like, and for each visual variable (see Tab.~\ref{tab:questions}). 
	
	We, additionally, performed one experiment to verify the visual distinguishability between the three shape type, sinusoidal, rectangular and sawtooth-like for combinations of low frequencies / low amplitudes and high frequencies / high amplitudes not listed in Tab.~\ref{tab:questions}. The recognition rates were approx. 92\%, 99\% and 99\% for sinusoidal, rectangular and sawtooth-like, respectively.
\end{description}

The design of the experiment assumed to have ``cooperative'' participants, \ie participants that won't ``cheap their way through'' the experiment, and that the time limit for the ability to concentrate is at most 20-25~min.
Tab.~\ref{tab:questions} states the number of experiments taken per experiment type. Each participant was asked to go through 90 experiments in total. 
We did not perform an evaluation of the dependency between the geometric and color visual variables, as this would have required a significantly larger amount of participants, beyond the pool of ``cooperative'' participants we have access to.        

\begin{table*}[bt]
	\renewcommand{\arraystretch}{1.2}
	\centering
	\resizebox{\textwidth}{!}{\begin{tabular}{|c|c|c|c|c|c|c|}\hline
			\textbf{Experim.}& \textbf{Shape} & \textbf{ Amplitude} & \textbf{Frequency} &  \textbf{Col. Amplitude} &  \textbf{Col. Frequency} & \textbf{\# exp.} \\\hline
			\textbf{Ampl1} & sin. & $\mathbf{\left[2\lldots  10\right]\rightarrow\left[0\lldots  12\right]}$ & $F\left[6\right]$ & n.a. & n.a. & 6\\\hline
			\textbf{Ampl2} & sin. & $\mathbf{\left[2\lldots  10\right]\rightarrow\left[0\lldots  12\right]}$ & $R\left[2\lldots  10\right]$ & n.a. & n.a. & 20\\\hline
			\textbf{Freq1} & sin. & $F\left[6\right]$ & $\mathbf{\left[2\lldots  10\right]\rightarrow\left[0\lldots  12\right]}$ & n.a. & n.a. & 6\\\hline
			\textbf{Freq2} & sin. & $R\left[2\lldots  10\right]$ & $\mathbf{\left[2\lldots  10\right]\rightarrow\left[0\lldots  12\right]}$ & n.a. & n.a. & 20\\\hline
			\textbf{SawtAmpl} & sin.$\rightarrow$sawt. & $\mathbf{x_{\mid\in\left[2\lldots  10\right]}\rightarrow\left[x-2 \lldots x+2\right]}$ & $F\left[6\right]$ & n.a. & n.a. & 5\\\hline
			\textbf{RectAmpl} & sin.$\rightarrow$rect. & $\mathbf{x_{\mid\in\left[2\lldots  10\right]}\rightarrow\left[x-2 \lldots x+2\right]}$ & $F\left[6\right]$ & n.a. & n.a. & 5\\\hline
			\textbf{SawtFreq} & sin.$\rightarrow$sawt. & $F\left[6\right]$ & $\mathbf{x_{\mid\in\left[2\lldots  10\right]}\rightarrow\left[x-2 \lldots x+2\right]}$ & n.a. & n.a. & 5\\\hline
			\textbf{RectFreq} & sin.$\rightarrow$rect. & $F\left[6\right]$ & $\mathbf{x_{\mid\in\left[2\lldots  10\right]}\rightarrow\left[x-2 \lldots x+2\right]}$ & n.a. & n.a. & 5\\\hline
			\textbf{ColAmpl1} & n.a. & n.a. & n.a. & $\mathbf{\left[1\lldots 5\right]\rightarrow\left[0\lldots 5\right]}$ & $F\left[3\right]$  & 3\\\hline
			\textbf{ColAmpl2} & n.a. & n.a. & n.a. & $\mathbf{\left[1\lldots 5\right]\rightarrow\left[0\lldots 5\right]}$ & $R\left[1\lldots  5\right]$  & 6\\\hline
			\textbf{ColFreq1} & n.a. & n.a. & n.a. & $F\left[3\right]$ & $\mathbf{\left[1\lldots 5\right]\rightarrow\left[0\lldots 5\right]}$  & 3\\\hline
			\textbf{ColFreq2} & n.a. & n.a. & n.a. & $R\left[1\lldots  5\right]$ & $\mathbf{\left[1\lldots 5\right]\rightarrow\left[0\lldots 5\right]}$  & 6\\\hline
	\end{tabular}}
	\caption{Overview of survey experiments. Each row is one experiment type, where the \textbf{bold-faced} are the visual variables, \ie the perceptional parameters to be assessed. Other parameters might be fixed values, indicated by \emph{F}, or randomly selected, indicated as \emph{R}. The numbers given in []-brackets are the stimuli values defined in Tab.~\ref{tab:metric}. The individual experiments are of two kinds: $\mathbf{\left[2\lldots  10\right]\rightarrow\left[0\lldots  12\right]}$, for example, generates stimuli in the value range $\mathbf{\left[2\lldots  10\right]}$ and ask for assessing the perceptional values in $\mathbf{\left[0\lldots 12\right]}$, while $\mathbf{x_{\mid\in\left[2\lldots  10\right]}\rightarrow\left[x-2 \lldots x+2\right]}$ generates stimuli values $x$ in the range $\mathbf{\left[2\lldots  10\right]}$ and ask for assessing the perceptional values in the dependent range $\mathbf{\left[x-2 \lldots x+2\right]}$. }
	\label{tab:questions}
\end{table*}

\subsection{Survey Evaluation}
\label{s:model.survey}

We invited students and researchers mainly from our university from the fields of computer scientists and sociology to participate in our online survey, and an anonymous group of 73 persons participated. The average time to take the survey was $\approx 26$~min.

\subsubsection{Procedure}
\label{s:model.survey.proc}
Given the raw results from the survey experiments conducted by the participants, we determined the required stimulus-to-perception transformation, quantization and calibration parameters after having applied an outlier removal.
\paragraph*{Outlier removal.}	
First, the ``senseless'' answers are filtered out, \ie answers which deviate from the expected value to an extend not explainable by the subjective character of perception alone. These outliers are mainly caused, \eg by a misunderstanding of the respective experimental setting or by an external distraction of the participant while conducting the experiment. We apply the two-step Chebyshev outlier detection method of Amidan~\etal~\cite{Amidan2005OutlierChebyshev}, with the filtering parameters $p_1 = 0.375$ and $p_2 = 0.175$ for all visual variables.

\paragraph*{Modeling the stimulus-to-perception transformation function.}	
Following Stevens~\cite{Stevens1957PsychophysLaw}, we assume that the stimulus-to-perception transformation has the form of a power function, \ie $g(\mathbf{x})=a\cdot\mathbf{x}^b + c$. Thus, having the perceived magnitudes, as stated by the participants, as data points $g(\mathbf{x})$ and the stimulus magnitudes $\mathbf{x}$ as the independent parameter, $a$, $b$ and $c$ are estimated using non-linear least-square fitting.

\paragraph*{Quantization.}
The aim is to find a quantization step such that all resulting magnitude levels do not overlap with neighboring confidence intervals for a given confidence level. This is analogous to the principle applied by estimation of just noticeable difference (jnd), which is also defined regarding the probability of correct assignments, usually 50\%, which we also apply in our experiment. Tab.~\ref{tab:quant} gives an overview of the quantization steps and the resulting number of discrete levels for each visual variable..

More precisely, the quantization step is calculated as follows:
\begin{compactenum}
	\item For each discrete stimulus magnitude level, observe and model the distribution of perceived magnitudes,
	\item Compute the 50\% confidence interval, symmetrically placed about the respective mean.
	\item Use the largest confidence interval as quantization step.
\end{compactenum}

\paragraph*{Shape-dependent calibration.}
We aim to define a function that calibrates the stimulus amplitude and frequency of a rectangular or sawtooth-like shape against the respective parameters of a sinusoidal shape. This calibration is modeled as a linear function $h(\mathbf{x})=k\cdot\mathbf{x}+l$, with $\mathbf{x}$ being the reference sinusoidal magnitudes from the calibration experiments (see Sec.~\ref{s:model.design}) and the perceived magnitudes regarding the rectangular or sawtooth-like shape as data points. The fitting of $h$ is done with a least-square method.

\subsubsection{Results}
\label{s:model.survey.res}
\begin{table}[tb]
	\centering
	\tiny 
	\resizebox{0.8\textwidth}{!}{\begin{tabular}{|c|c|c|c|c|}\hline
			\textbf{Vis. Variable} & Shape Ampl. & 	Shape Freq. & Col. Ampl. & Col Freq.\\\hline
			\textbf{\# data} &1796 & 1787  & 648 & 646\\\hline
			\textbf{\# outliers} & 102 & 111 & 9 & 11 \\\hline	
	\end{tabular}}
	\caption{Outlier removal results.} 
	\label{tab:outliers}
\end{table}

\begin{figure}[h!]
	\centering
	\subfloat[]{
		\includegraphics[width=0.24\textwidth]{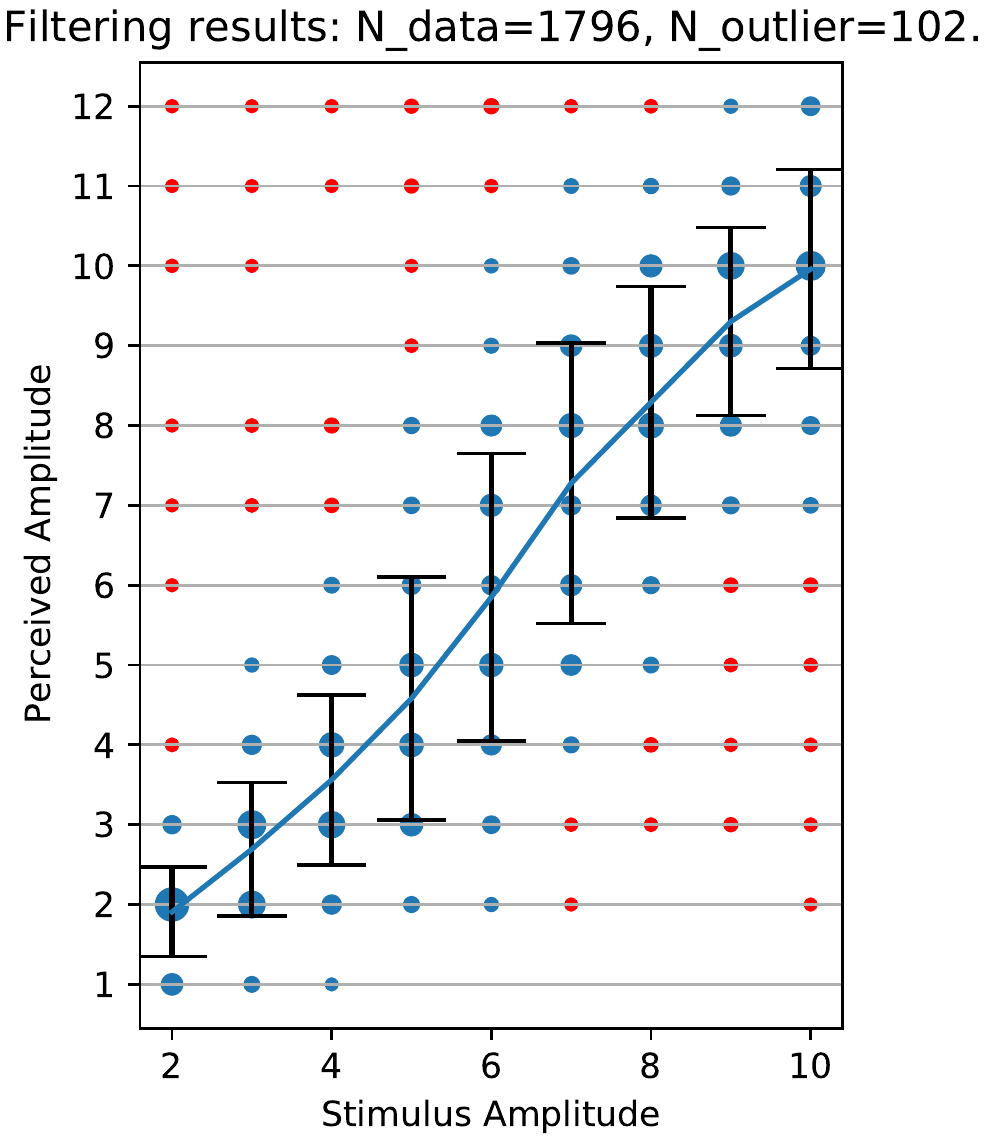} 
		\label{fig:amplFilt}
		
	}
	\subfloat[]{
		\includegraphics[width=0.24\textwidth]{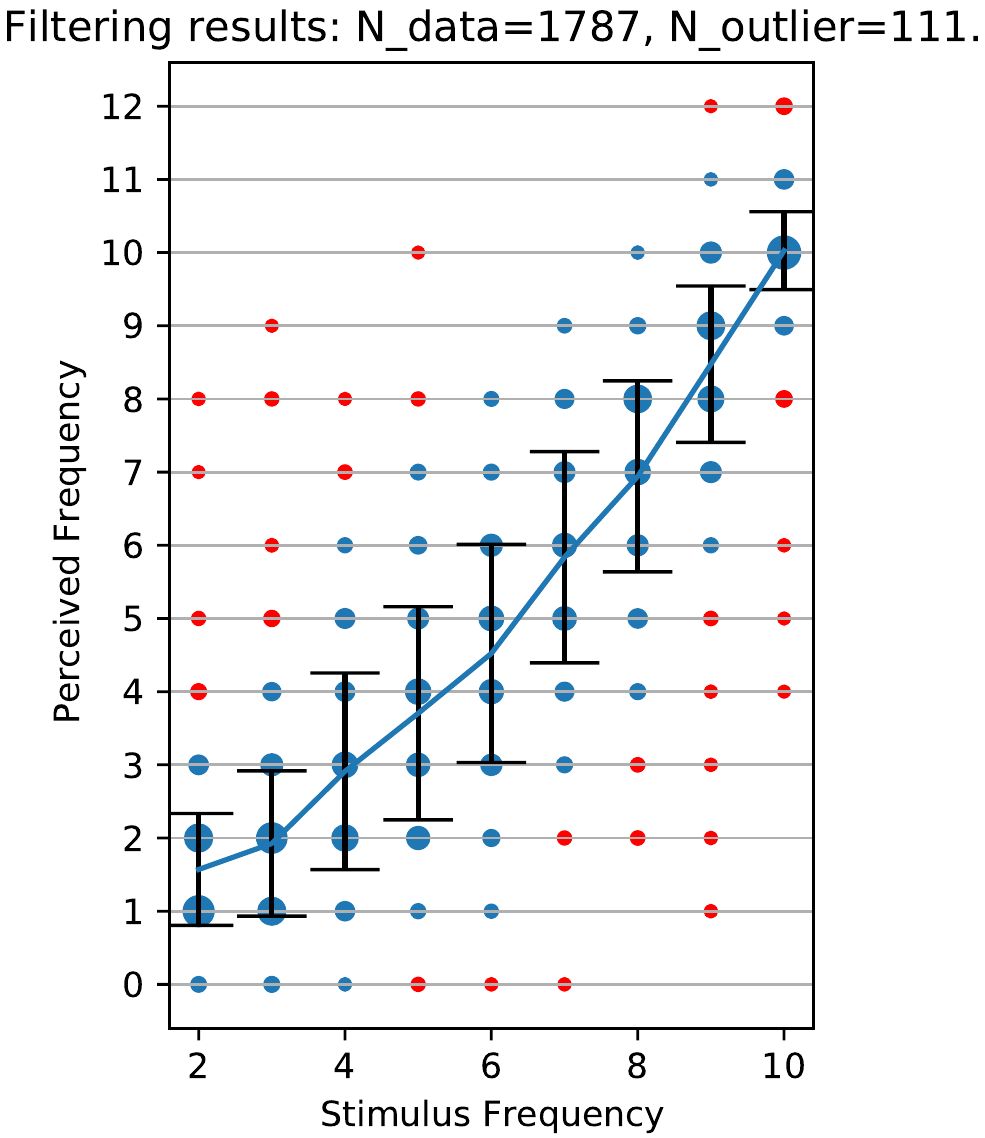}
		\label{fig:freqFilt}
	}
	\subfloat[]{
		\includegraphics[width=0.24\textwidth]{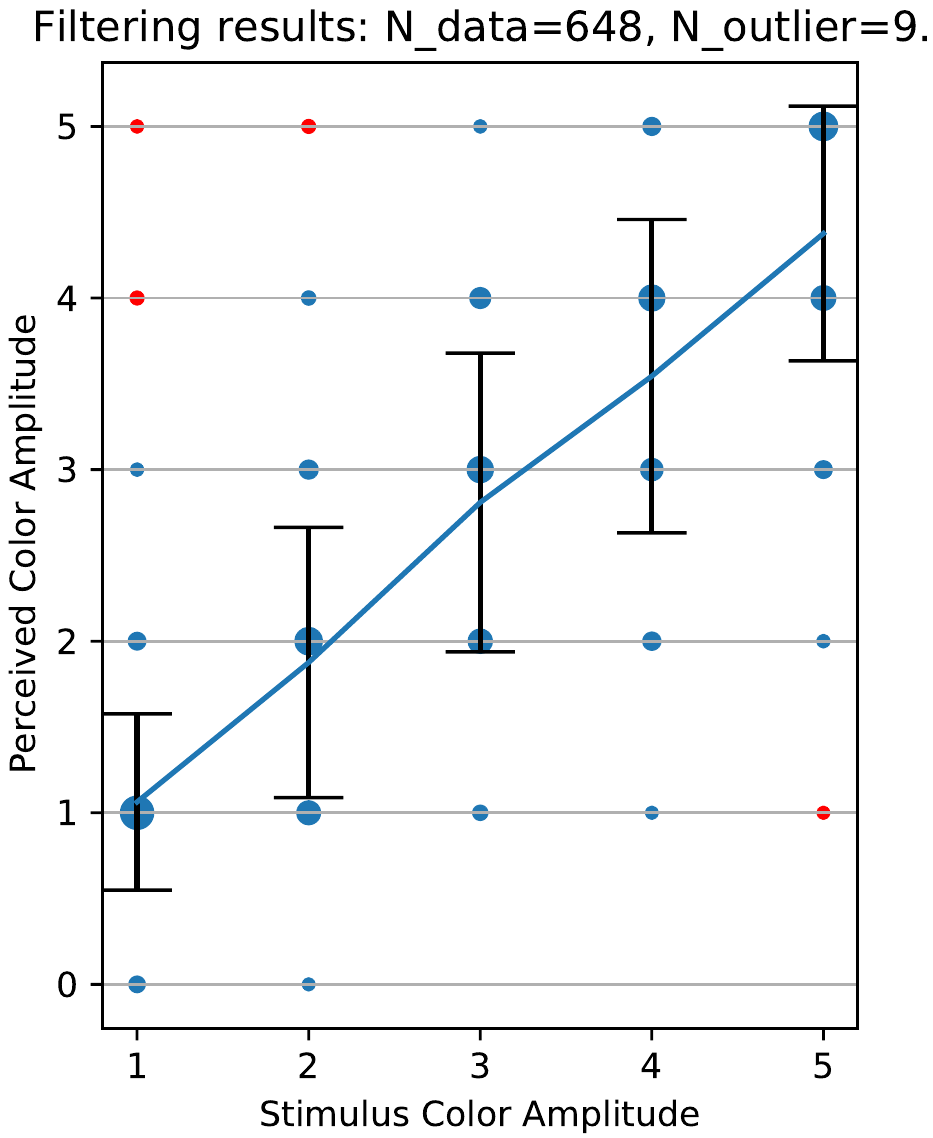} 
		\label{fig:colAmplFilt}
		
	}
	\subfloat[]{
		\includegraphics[width=0.24\textwidth]{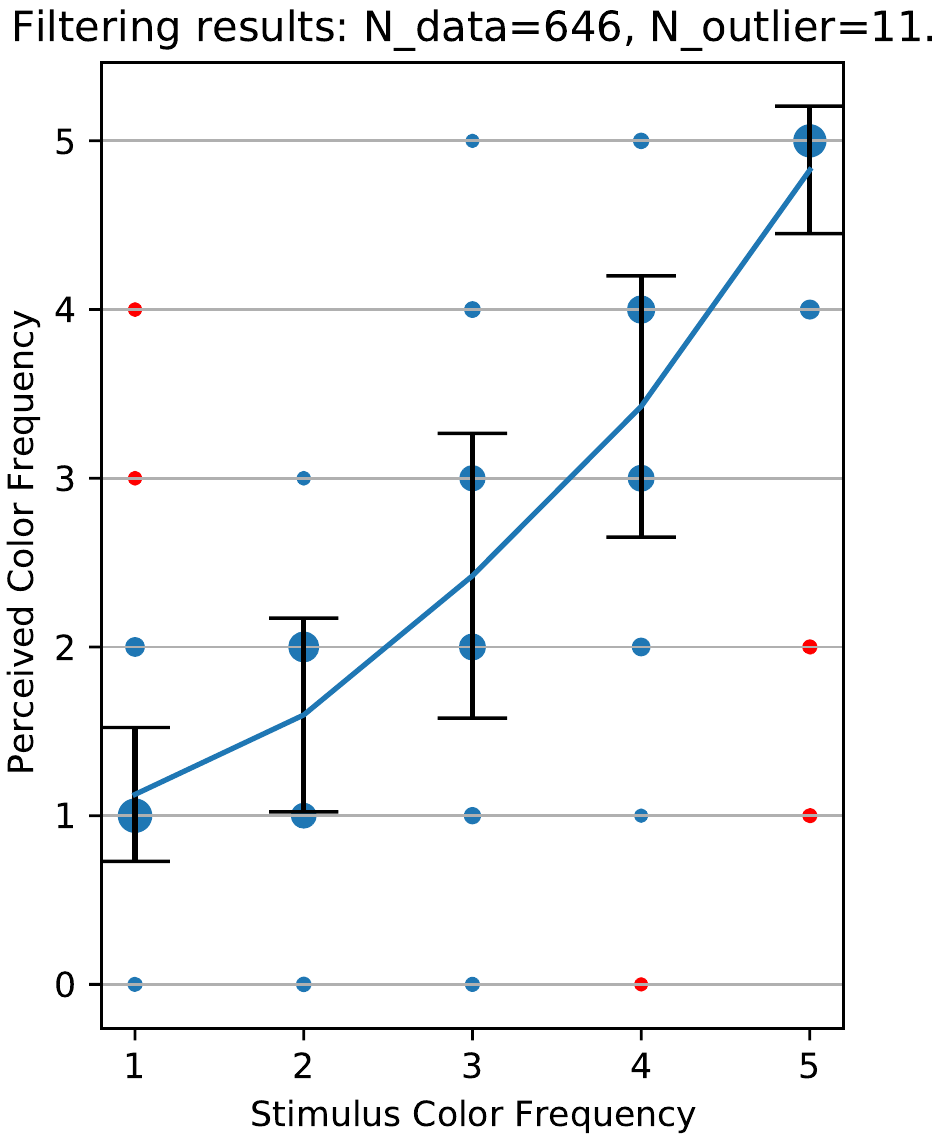}
		\label{fig:colFreqFilt}
	}
	
	\caption{Results of the two-step Chebyshev Outlier Detection: outliers are in red; the point size encodes the number of occurrences.}
	
	\label{fig:filteredData}
\end{figure}

\paragraph*{Outlier removal.}
Tab.~\ref{tab:outliers} gives an overview for all visual variables. The overall amount of removed outlier is 4.8\%. The detailed statistics of the outlier removal are represented in Fig.~\ref{fig:filteredData}.

\begin{figure}[h!]
	\centering
	\subfloat[]{
		\includegraphics[width=0.75\textwidth]{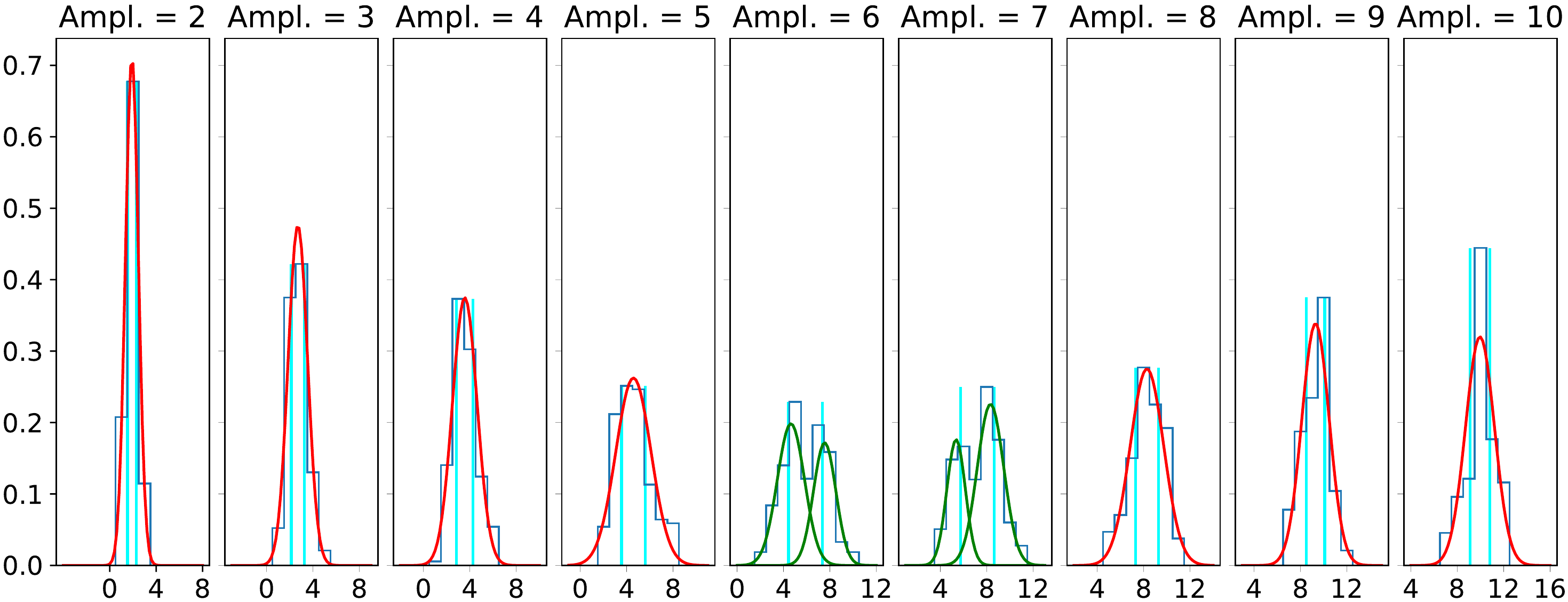} 
	}\\	
	\subfloat[]{
		\includegraphics[width=0.75\textwidth]{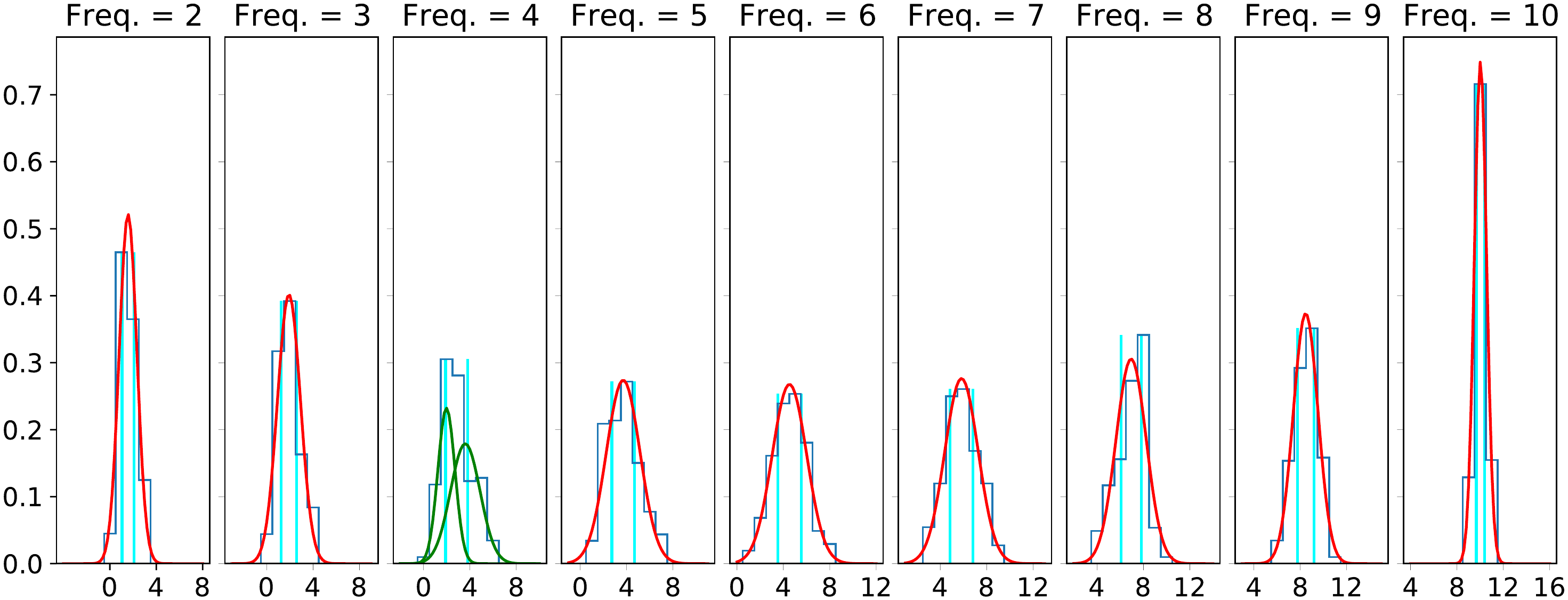} 
	}\\	
	\subfloat[]{
		\includegraphics[width=0.75\textwidth]{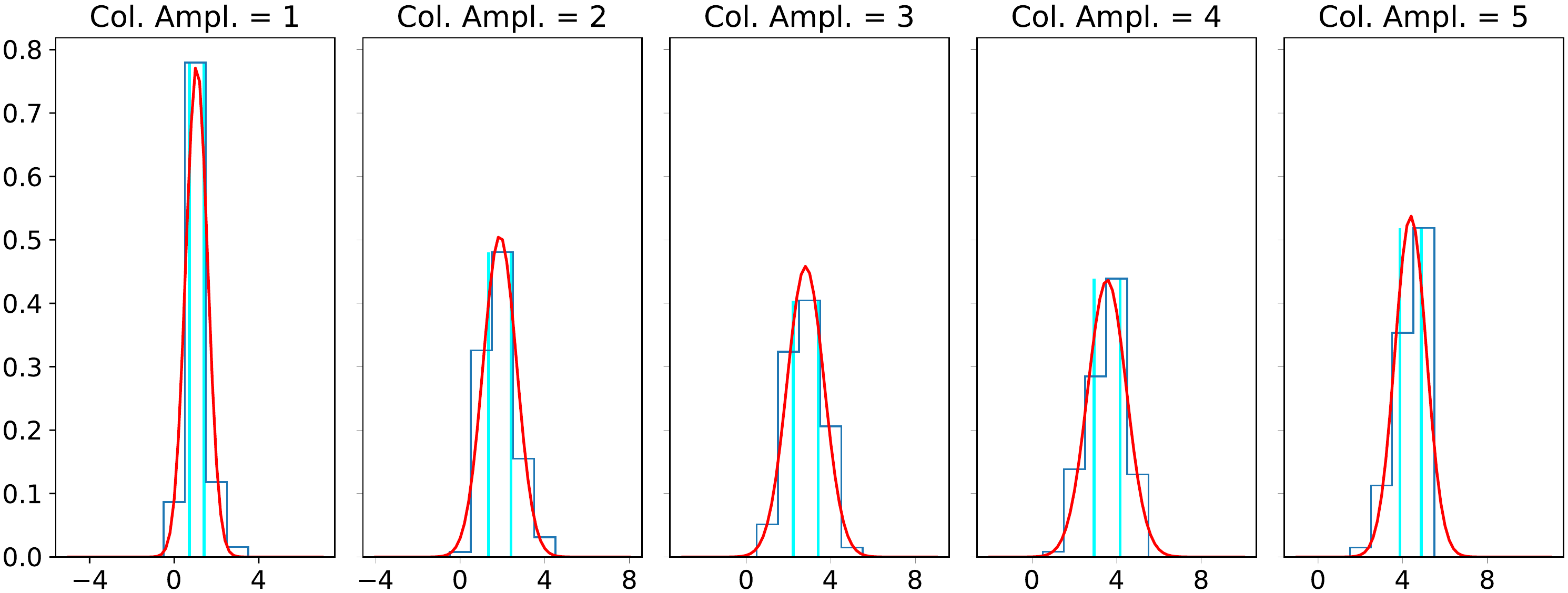} 
	}\\	
	\subfloat[]{
		\includegraphics[width=0.75\textwidth]{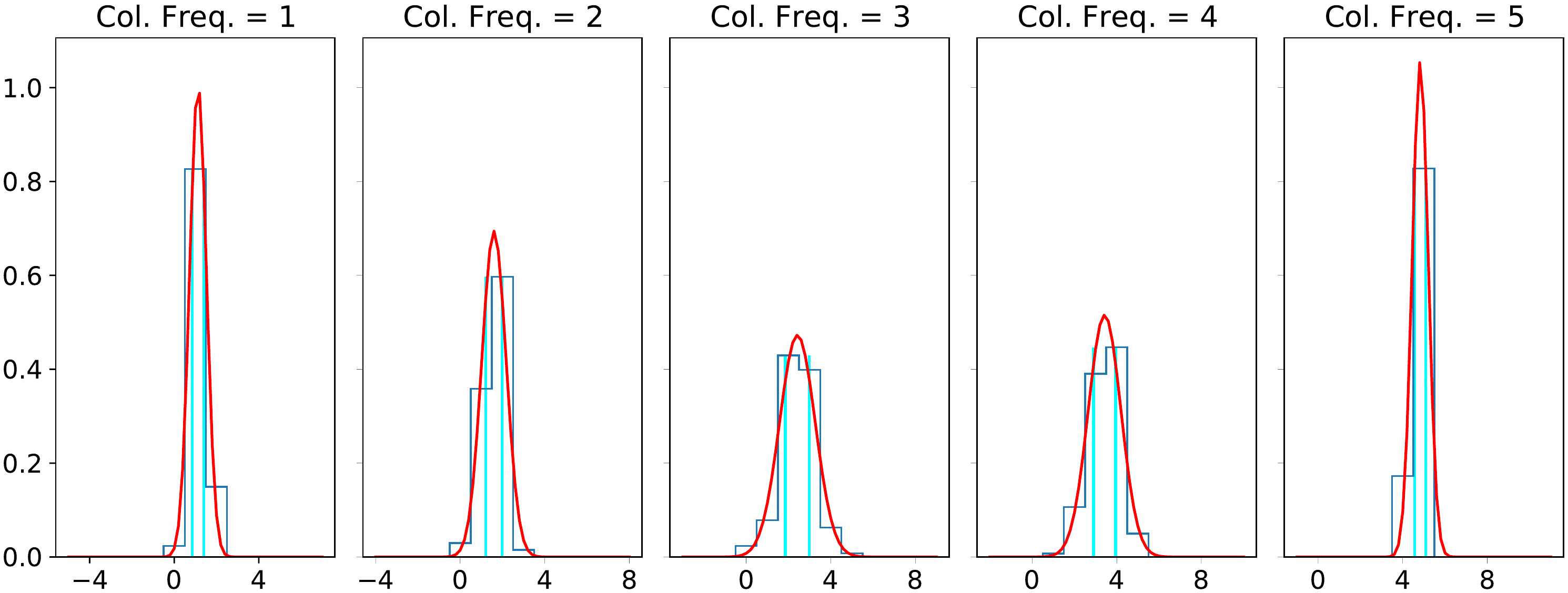} 
	}	
	\caption{Perceived magnitudes as normal distributions. Red: simple Gaussian; green: two-component Gaussian mixture; cyan: borders of 50\% confidence intervals.}
	
	\label{fig:errorDist}
\end{figure}

\begin{figure}[bt]
	\centering
	\subfloat[]{
		\includegraphics[width=0.24\textwidth]{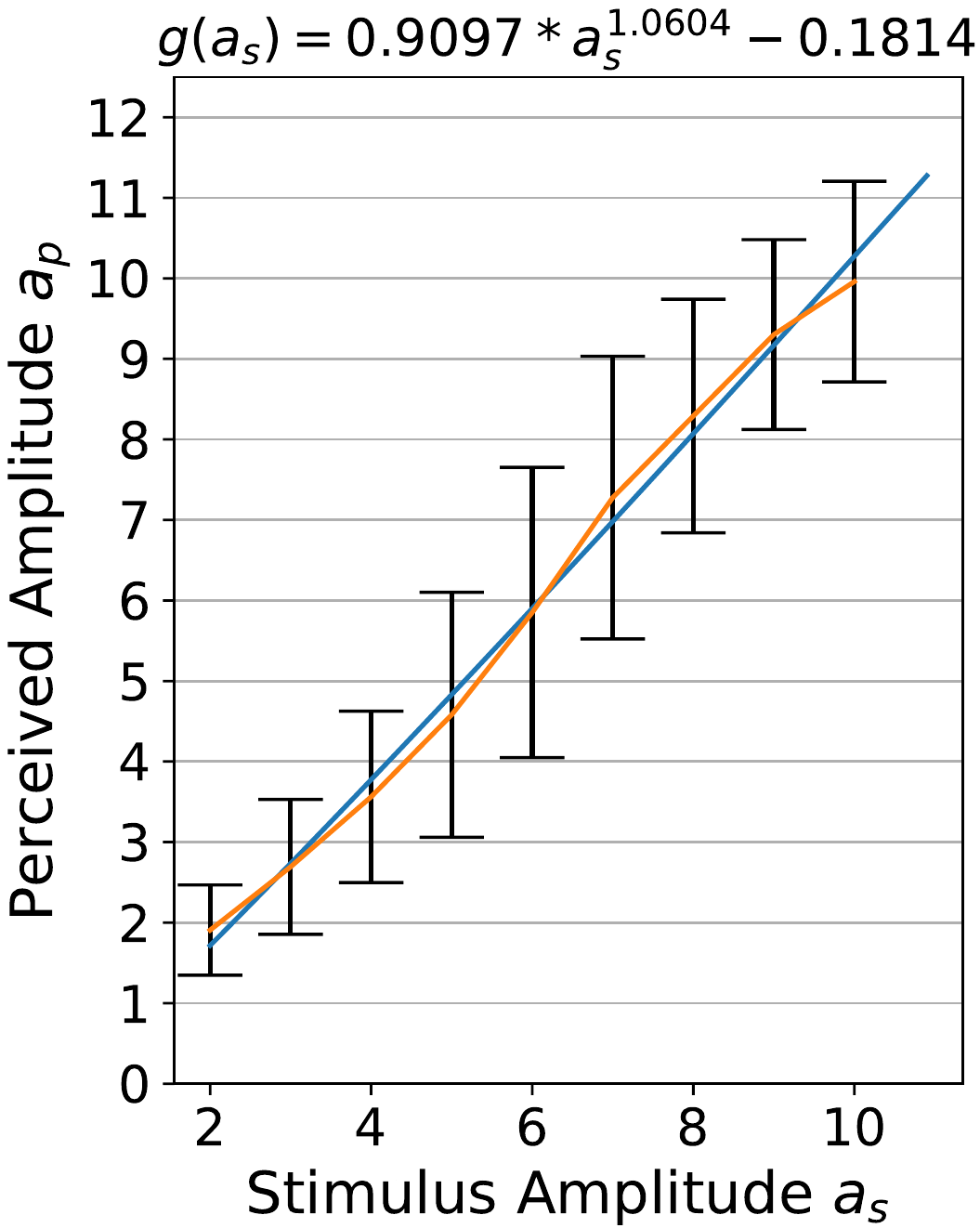} 
		\label{fig:amplFitting}
		
	}
	\subfloat[]{
		\includegraphics[width=0.24\textwidth]{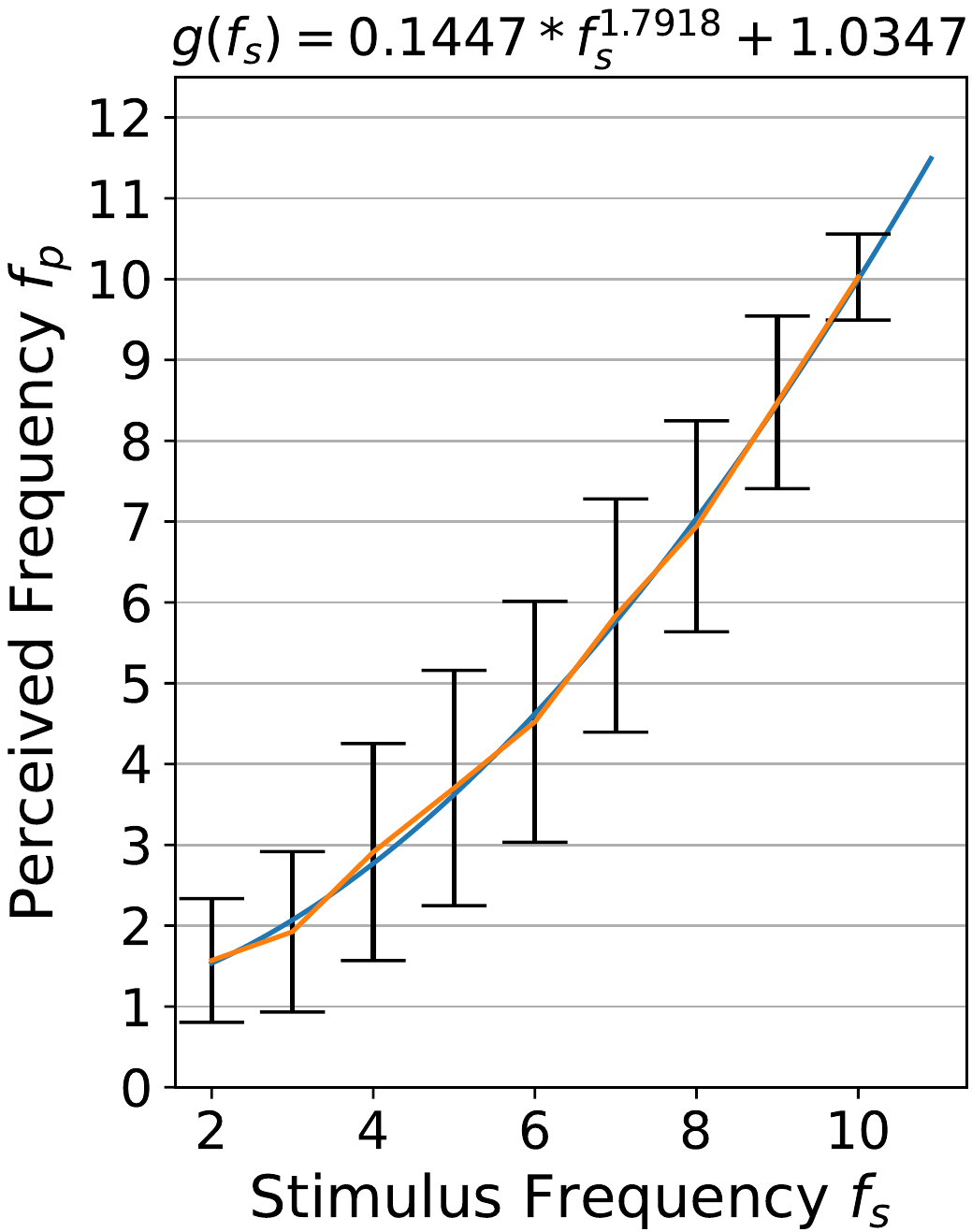} 
		\label{fig:freqFitting}
		
	}
	\subfloat[]{
		\includegraphics[width=0.24\textwidth]{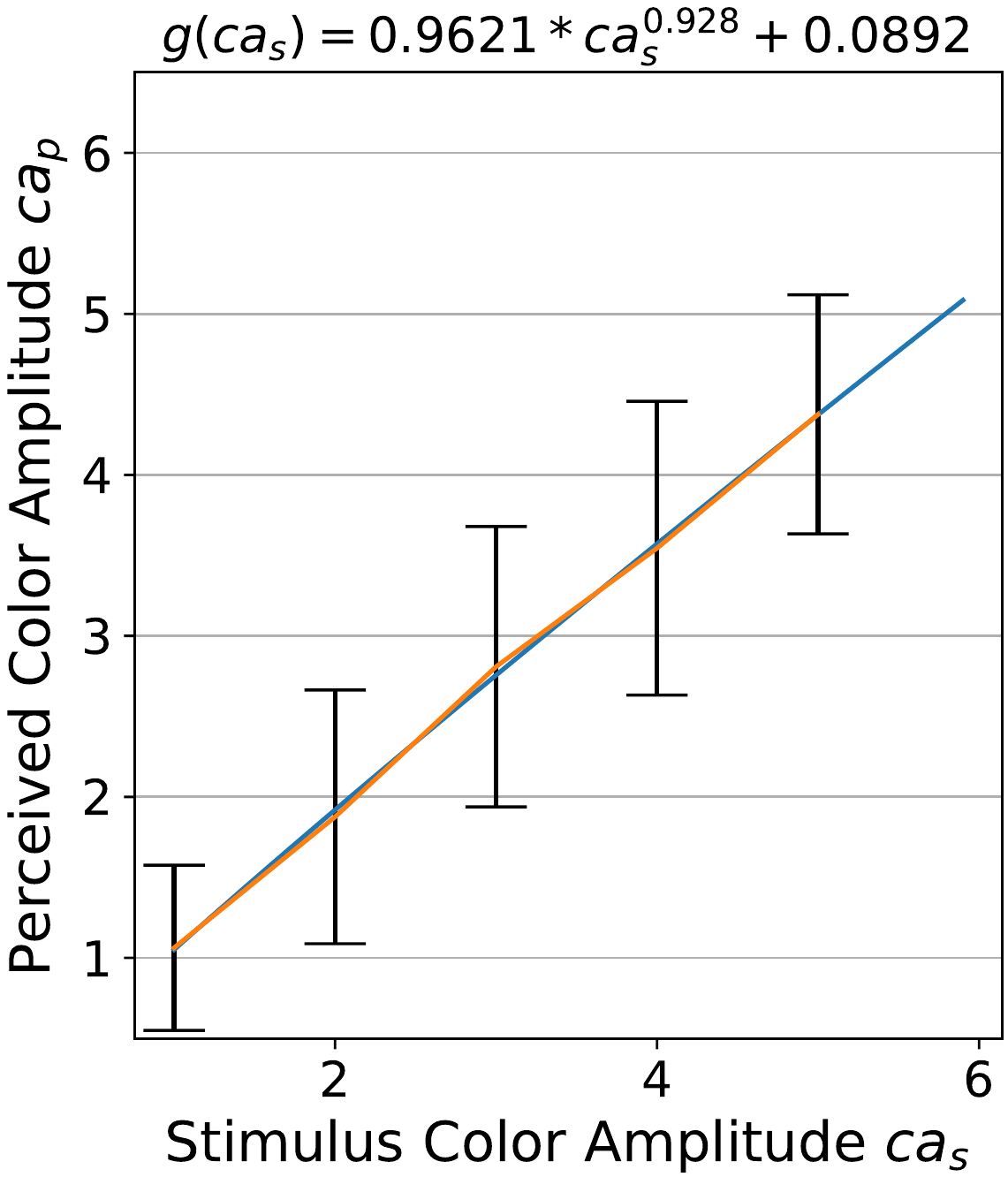} 
		\label{fig:colAmplFitting}
		
	}
	\subfloat[]{
		\includegraphics[width=0.24\textwidth]{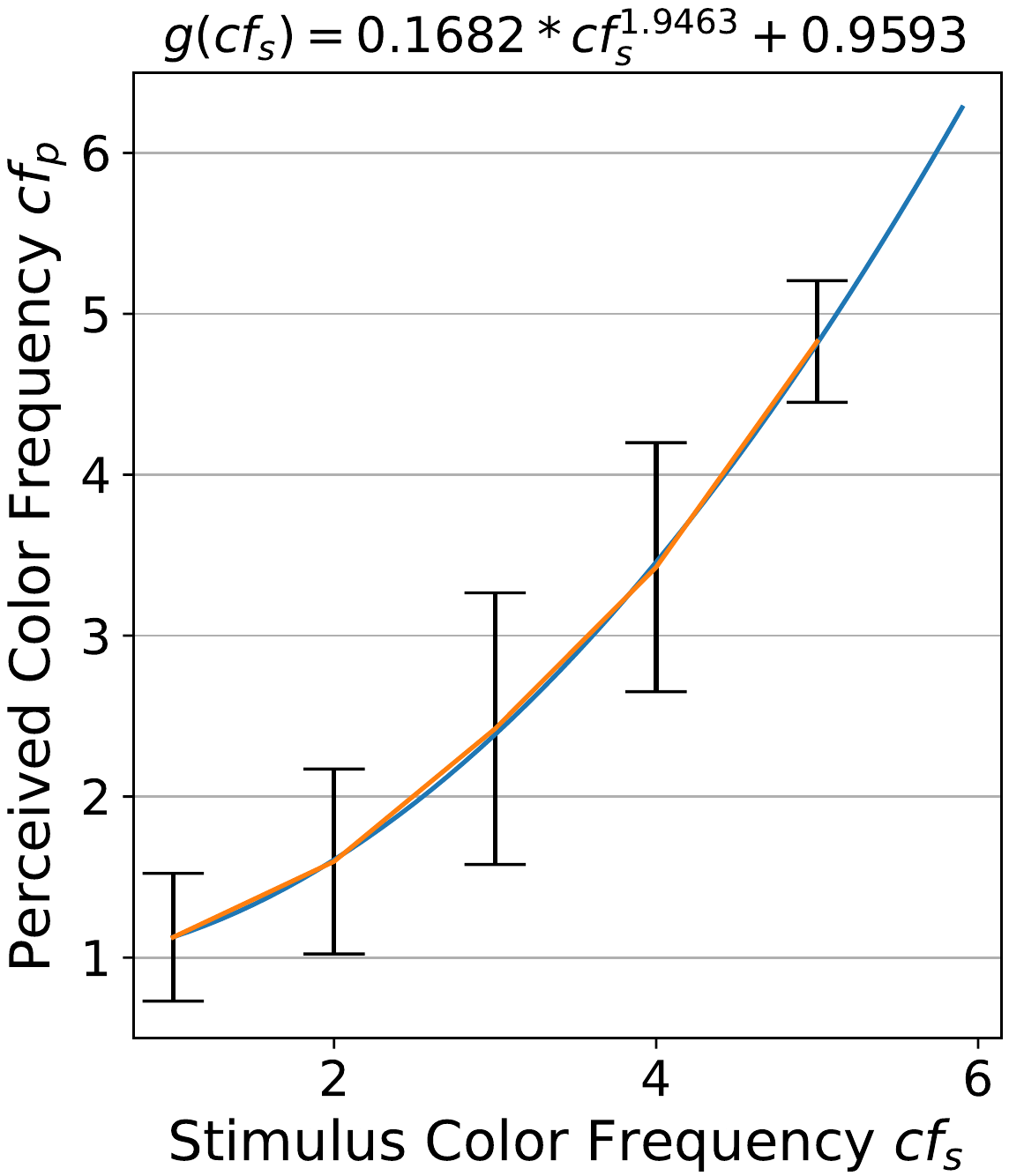} 
		\label{fig:colFreqFitting}		
	}
	\caption{Modeling stimulus-to-perception transformation: blue: fitted transformation functions; orange: lines connecting perceptual means.}
	
	\label{fig:transfFitting}
\end{figure}

\paragraph*{Stimulus-to-perception transformation function.}
The modeled functions for transformation of stimuli into perceived magnitudes are presented in Fig.~\ref{fig:transfFitting}. For both, geometry and color, we observe a rather linear and positive power dependency of the perceived magnitudes from the stimulus magnitudes for amplitude and frequency, respectively.

\paragraph*{Quantization}
The distributions of the perceived magnitudes reveal a mono-modal Gaussian nature for most magnitude levels and a bi-modal Gaussian behavior for the shape amplitude and shape frequency for medium amplitude and frequency values. The latter can be explained by a larger distance to the min. and max. references, which can be seen as a design-related phenomenon. Fig.~\ref{fig:errorDist} shows the distribution of the perceived magnitudes with the respective 50\% confidence intervals for each visual variable.
We compute the 50\% confidence for a bi-modal Gaussian distribution by identifying the 25\% and 75\% limits of the cumulative distribution of the superposition of both Gaussians.

Tab.~\ref{tab:quant} gives an overview of the quantization steps and the resulting number of discrete levels for each visual variables. 
\begin{table}[bt]
	\centering
	\tiny 
	\resizebox{0.8\textwidth}{!}{\begin{tabular}{|c|c|c|c|c|}\hline
			\textbf{Vis. Variable} & Shape Ampl. & Shape Freq .& Col. Ampl. & Col Freq.\\\hline
			Quant. step $\Delta v$ (adu)& $\approx 2.91$ & $\approx 2.01$ & $\approx 1.23$ & $\approx 1.14$  \\\hline
			\# levels ($50$~mm)& 4 & 5 & 4 & 4  \\\hline	
	\end{tabular}}
	\caption{Quantization results: derived from the user experiment with the glyph's size $50\times50$~mm.}
	\label{tab:quant}
\end{table}

\begin{figure}[tb]
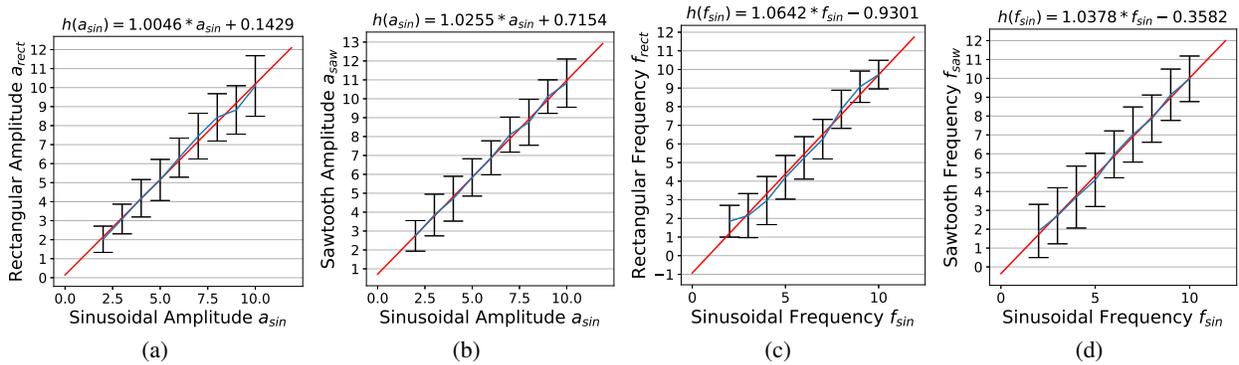

	\centering
	\subfloat[]{
		\includegraphics[width=0.24\textwidth]{amplCalibr_rect2} 
		\label{fig:amplCalibrRect}
		
	}
	\subfloat[]{
		\includegraphics[width=0.24\textwidth]{amplCalibr_saw2} 
		\label{fig:amplCalibrSaw}
		
	}
	\subfloat[]{
		\includegraphics[width=0.24\textwidth]{freqCalibr_rect2} 
		\label{fig:freqCalibrRect}
		
	}
	\subfloat[]{
		\includegraphics[width=0.24\textwidth]{freqCalibr_saw2} 
		\label{fig:freqCalibrSaw}		
	}
	\caption{Calibration of rectangular and sawtooth-like shape against sinusoidal shape.}		
	\label{fig:shapeCalibr}
\end{figure}

\paragraph*{Shape calibration}
Fig.~\ref{fig:shapeCalibr} shows the modeled linear functions for magnitude calibration against a sinusoidal reference, per shape and stimulus. The data demonstrate that the influence of a specific shape on the perception of amplitude and frequency magnitudes is rather marginal.

\section{GfI Examples}
\begin{figure}[h!]
	\captionsetup[subfloat]{farskip=2pt,captionskip=1pt}
	\centering
	\subfloat[original]{
		\includegraphics[width=0.245\textwidth]{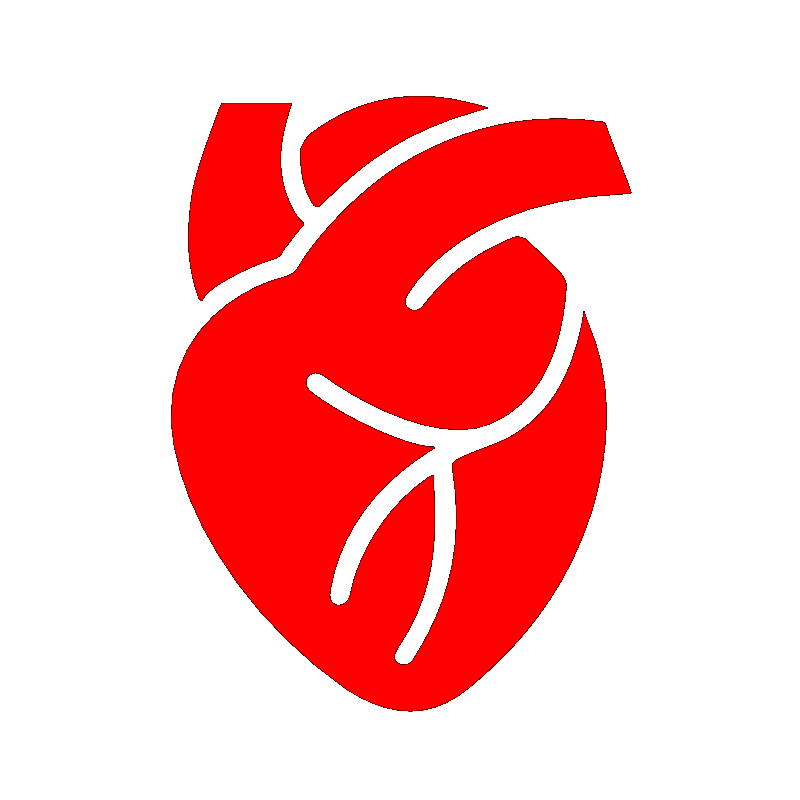} 
		\label{fig:heart.orig}
	}
	\subfloat[f=1, a=1]{
		\includegraphics[width=0.245\textwidth]{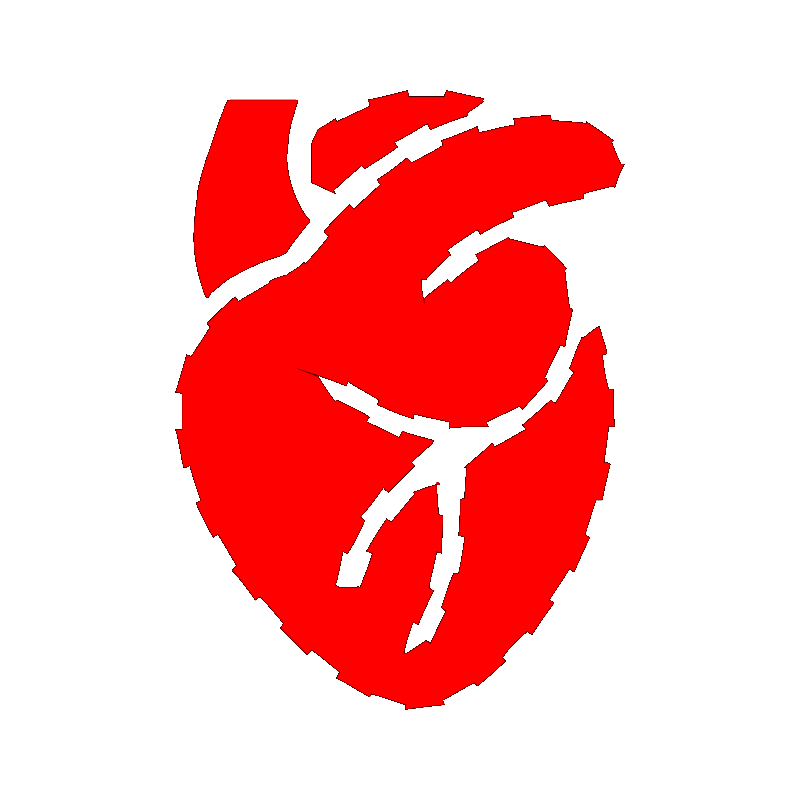} 
		\label{fig:heart.geo1}
	}	
	\subfloat[f=4, a=1]{
		\includegraphics[width=0.245\textwidth]{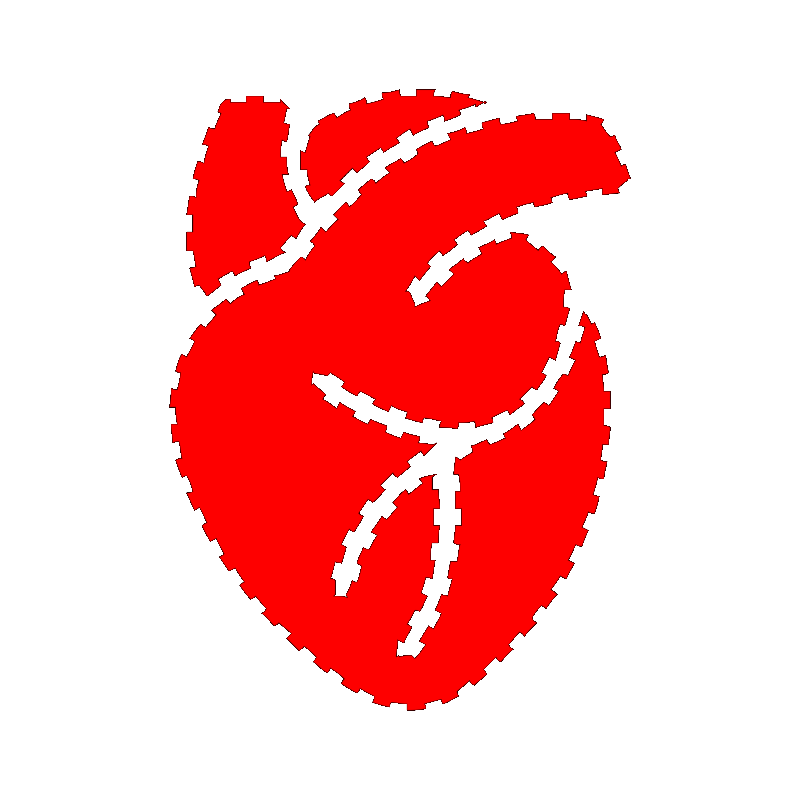} 
		\label{fig:heart.geo2}
	}	
	\subfloat[f=4, a=3]{
		\includegraphics[width=0.245\textwidth]{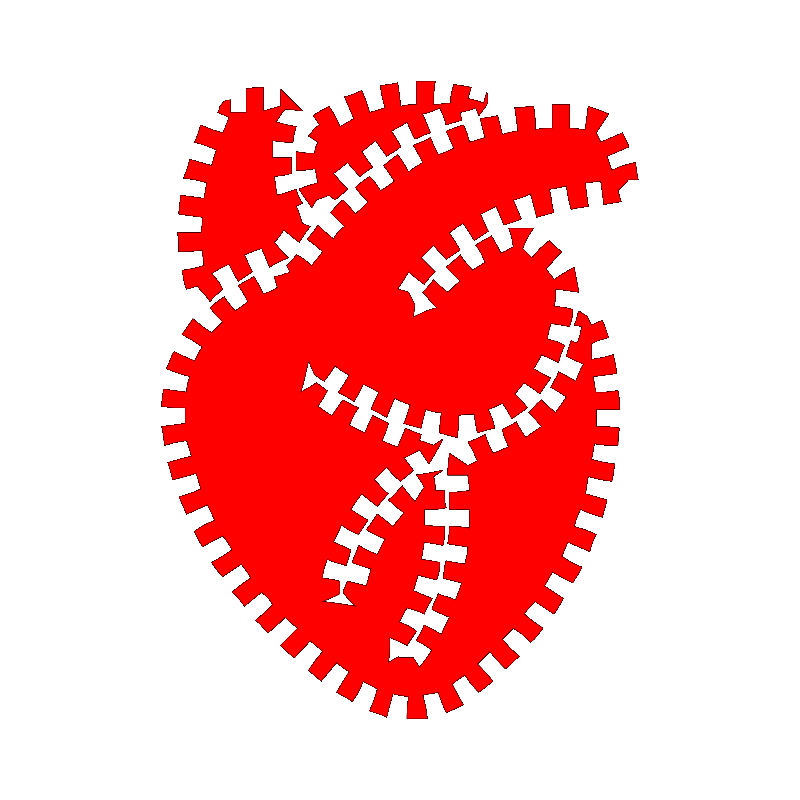} 
		\label{fig:heart.geo3}
	}\\
	\subfloat[cf=1, ca=1]{
		\includegraphics[width=0.245\textwidth]{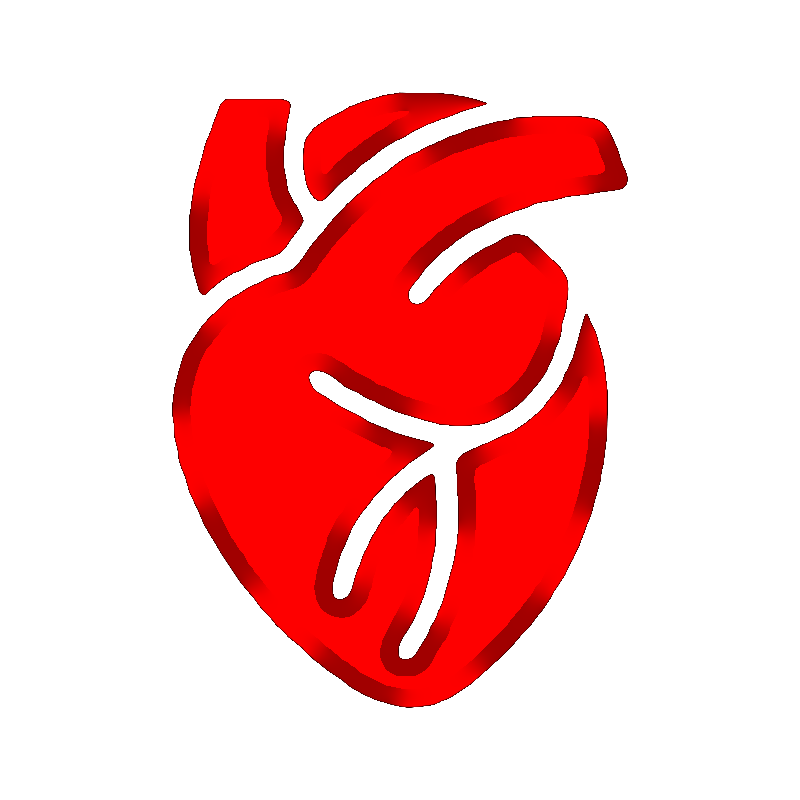} 
		\label{fig:heart.col1}
	}	
	\subfloat[cf=1, ca=3]{
		\includegraphics[width=0.245\textwidth]{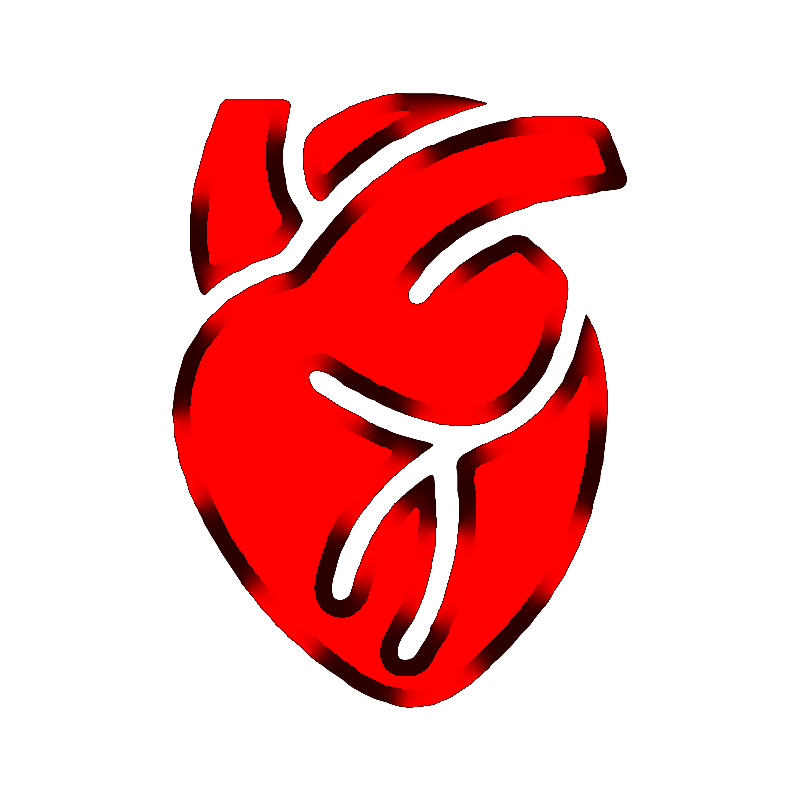} 
		\label{fig:heart.col2}
	}
	\subfloat[cf=4, ca=3]{
		\includegraphics[width=0.245\textwidth]{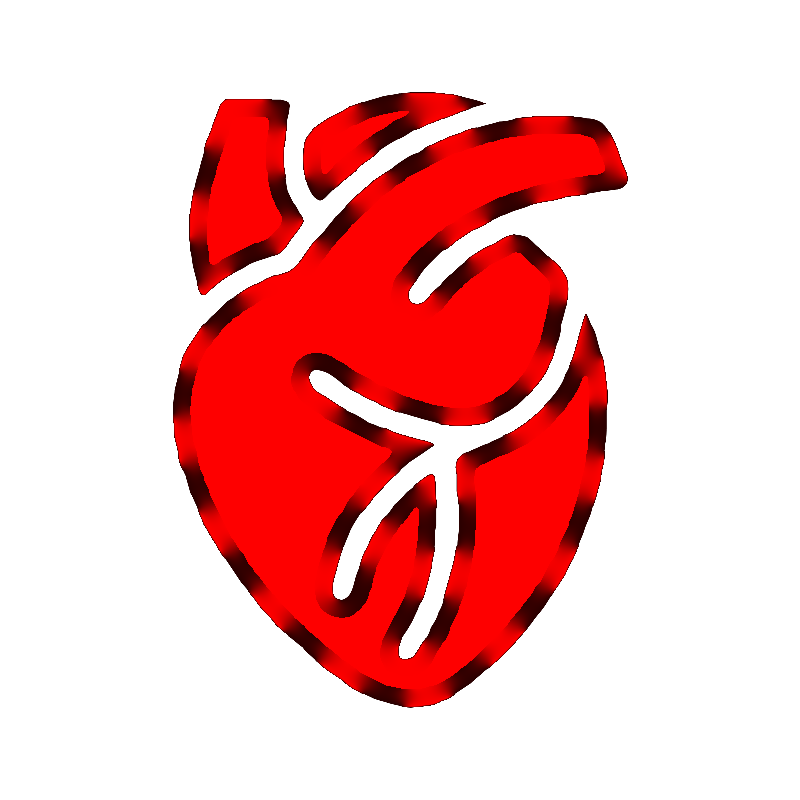} 
		\label{fig:heart.col3}
	}	
	\subfloat[f=4, a=3, cf=4, ca=3]{
		\includegraphics[width=0.245\textwidth]{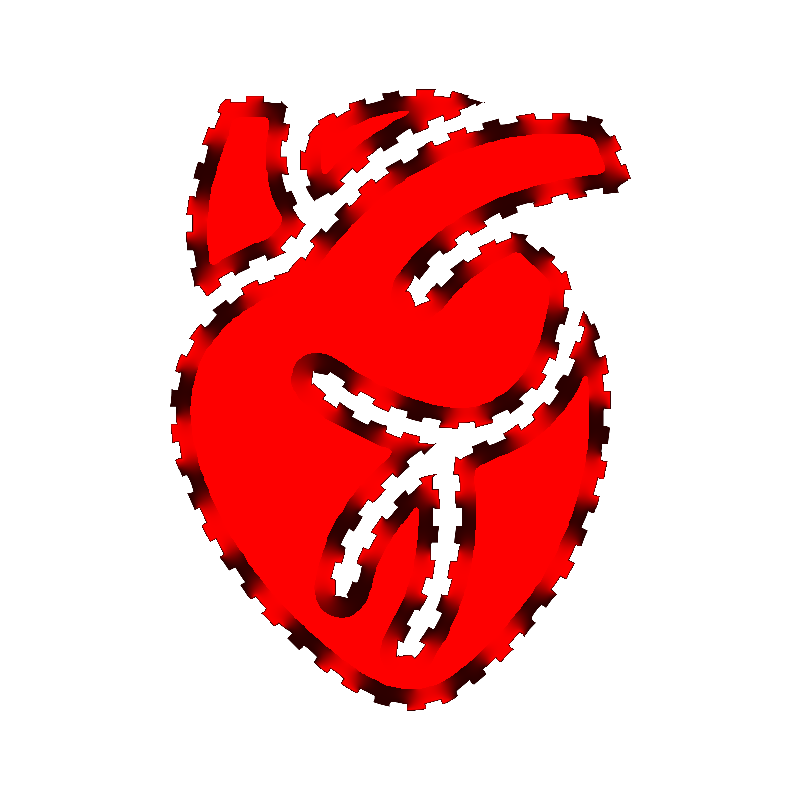} 
		\label{fig:heart.mix}
	}	
	\caption{Heart glyphs. Fig.~\ref{fig:heart.orig} shows the original icon. Figs.\ref{fig:heart.geo1}-\ref{fig:heart.geo3} and Figs.\ref{fig:heart.col1}-\ref{fig:heart.col3} present variations of geometric and color parameters, respectively. Fig.~\ref{fig:heart.mix} shows a combination of geometric and color visual variables. The left top part of the heart is not modified by the lowest frequency, \ie in Fig.~\ref{fig:heart.geo1}, because the corresponding curve is to short, having at the same time a relatively large number of control points, so that arc-length parametrization for the target frequency would require a knot deletion.}
	\label{fig:heart}	
\end{figure}

\begin{figure}[h!]
	\captionsetup[subfloat]{farskip=2pt,captionskip=1pt}
	\centering
	\subfloat[original]{
		\includegraphics[width=0.245\textwidth]{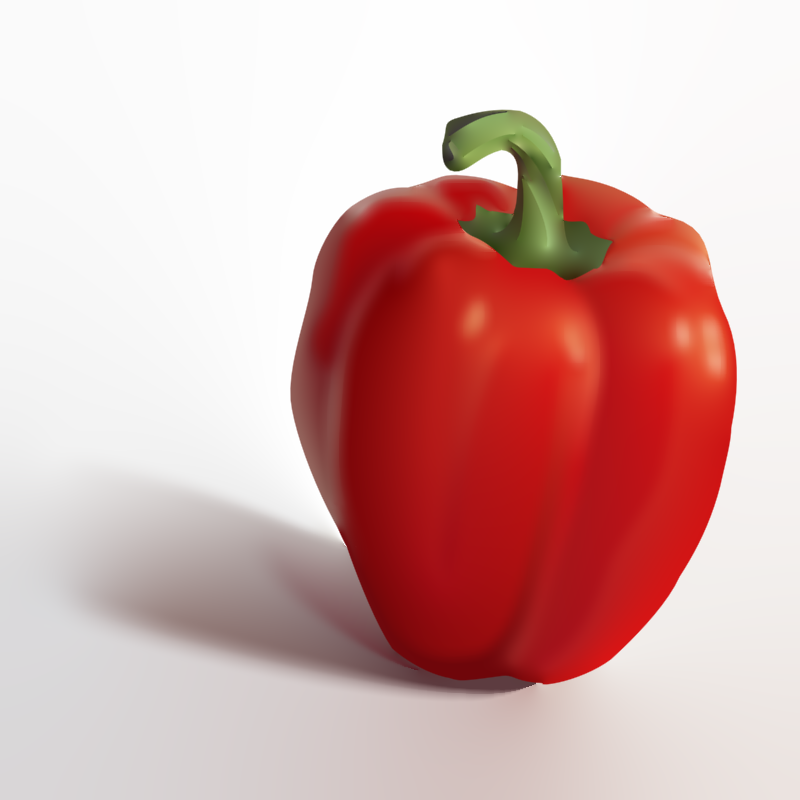} 
		\label{fig:pepper.orig}
	}
	\subfloat[s=rect., f=2, a=2]{
		\includegraphics[width=0.245\textwidth]{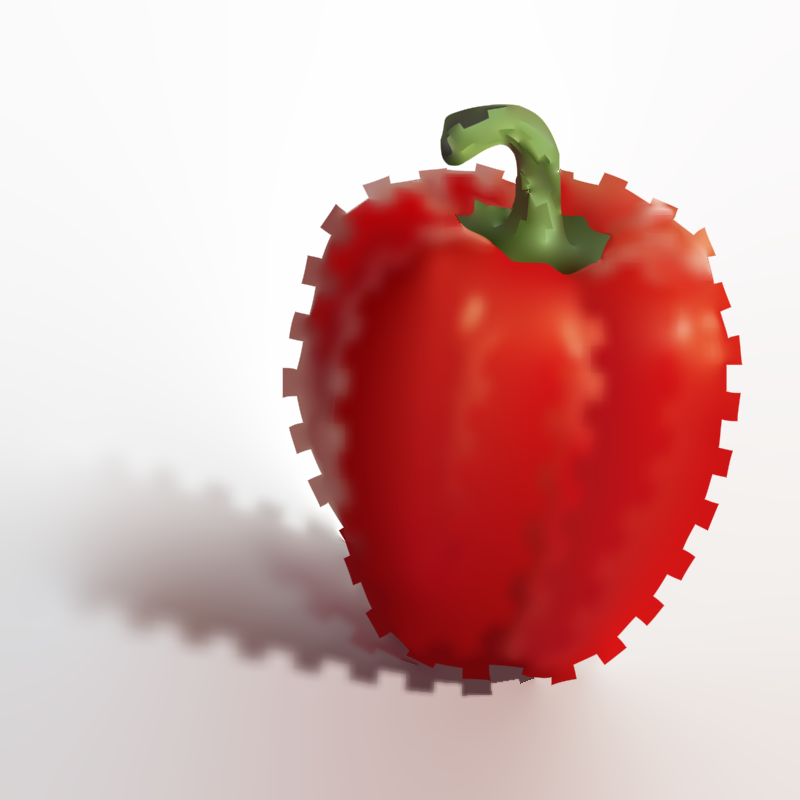} 
		\label{fig:pepper.rect}
	}	
	\subfloat[s=sin., f=2, a=2]{
		\includegraphics[width=0.245\textwidth]{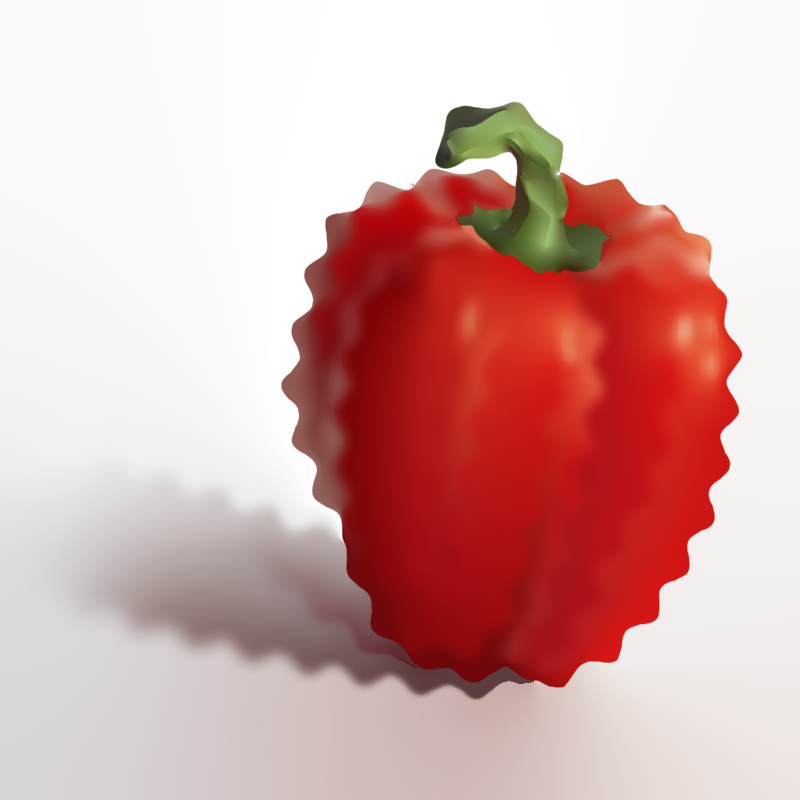} 
		\label{fig:pepper.sin}
	}	
	\subfloat[s=sawt., f=2, a=2]{
		\includegraphics[width=0.245\textwidth]{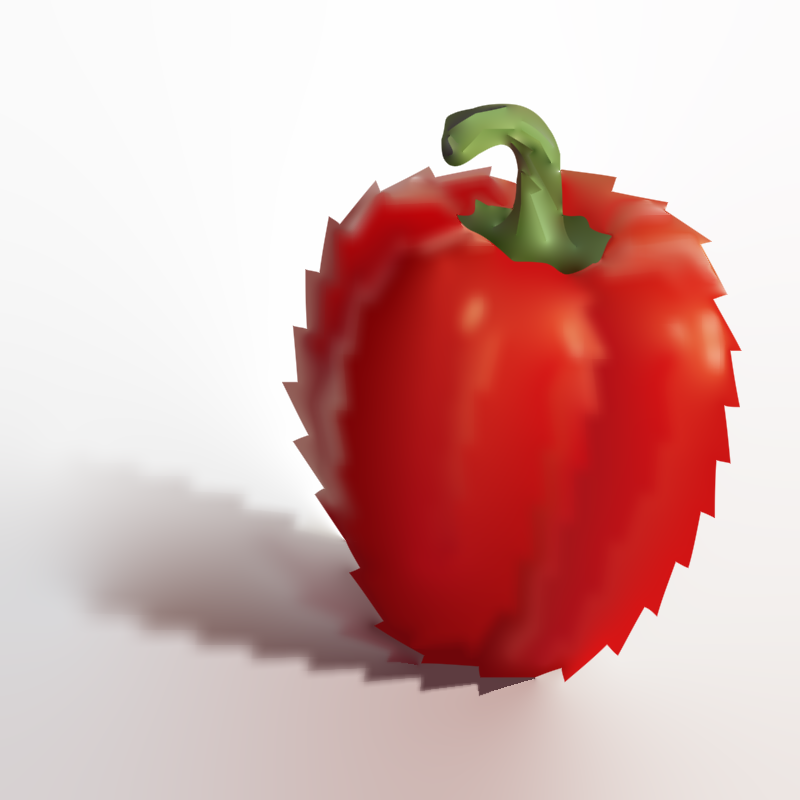}
		\label{fig:pepper.saw1}
	}\\
	\subfloat[s=sawt., f=4, a=2]{
		\includegraphics[width=0.245\textwidth]{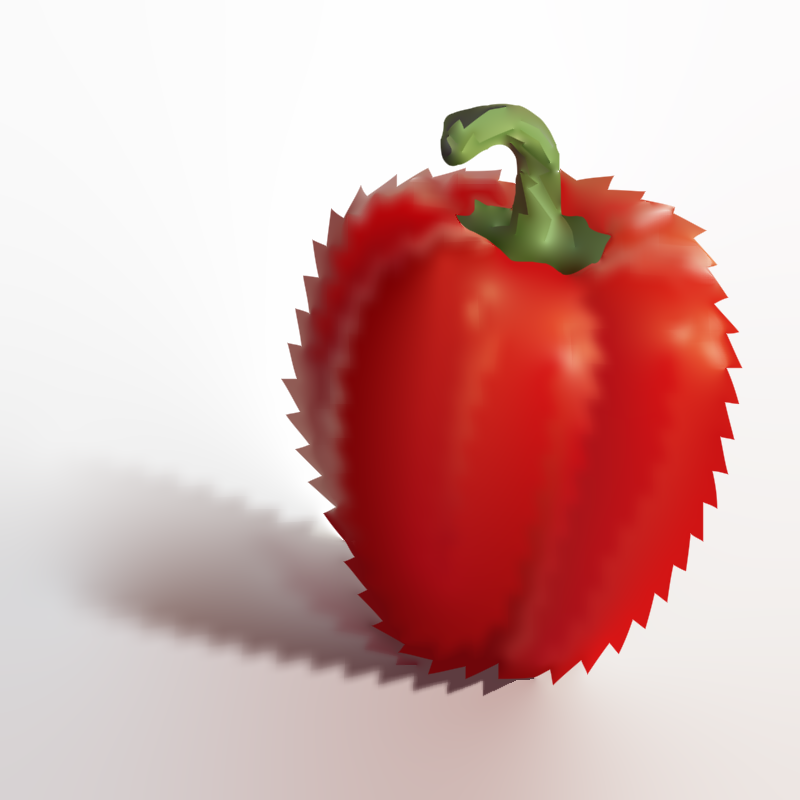}
		\label{fig:pepper.saw2} 
	}	
	\subfloat[s=sawt., f=4, a=3]{
		\includegraphics[width=0.245\textwidth]{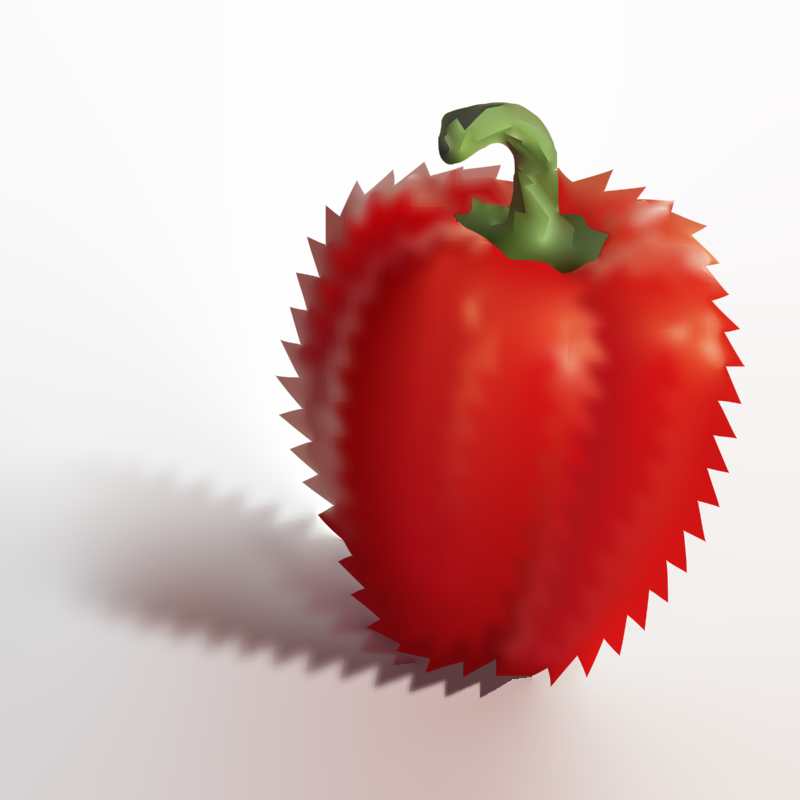}
		\label{fig:pepper.saw3} 
	}
	\subfloat[cf=3, ca=1]{
		\includegraphics[width=0.245\textwidth]{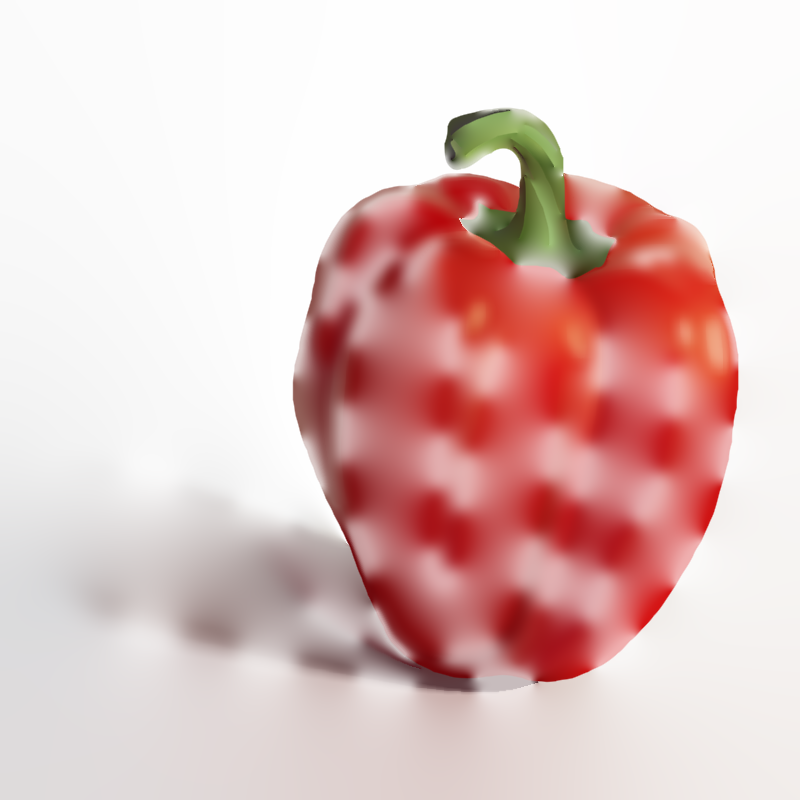} 
		\label{fig:pepper.col1}
	}	
	\subfloat[cf=4, ca=1]{
		\includegraphics[width=0.245\textwidth]{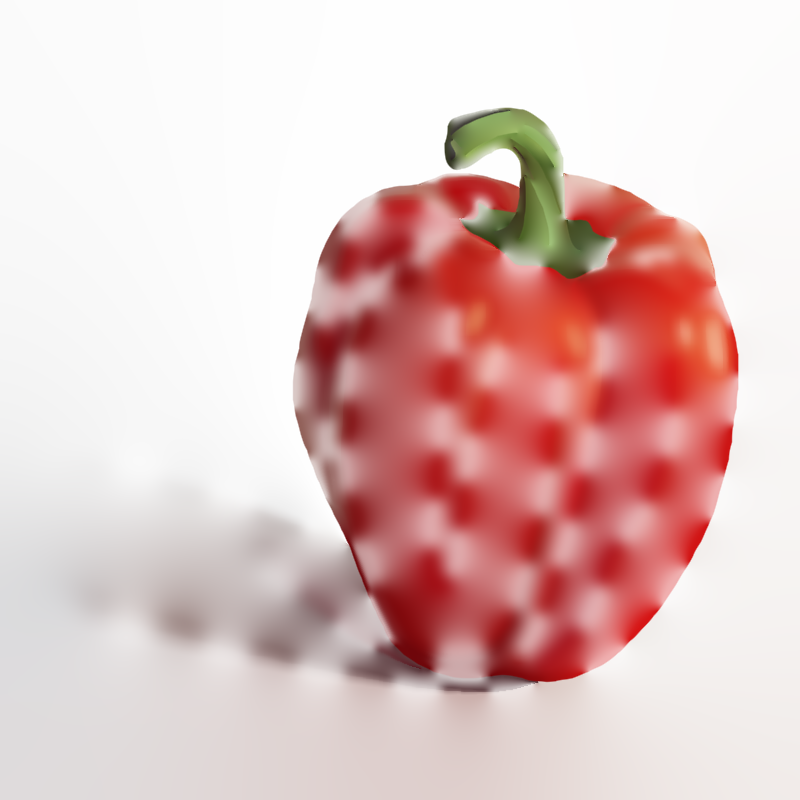} 
		\label{fig:pepper.col2}
	}	
	\caption{Pepper glyphs. Fig.~\ref{fig:pepper.orig} shows the original image. Figs.\ref{fig:pepper.rect}-\ref{fig:pepper.saw1} demonstrate construction of different shapes. Figs.\ref{fig:pepper.saw1}-\ref{fig:pepper.saw3} show variations of geometric frequency and amplitude and Figs.\ref{fig:pepper.col1}-\ref{fig:pepper.col2} represent color modifications with varying color frequency.} 
	\label{fig:pepper}	
\end{figure}

This section presents some examples of GfIs. Fig.~\ref{fig:heart} show seven glyphs with varying visual variables, constructed from a heart icon. Fig.~\ref{fig:pepper} demonstrates the results of our approach having a more complex pepper image as input with none-homogeneous color distribution. The visual variables with their respective levels, applied in the corresponding glyphs, are indicated in the figure descriptions as follows: \emph{s} for shape, \emph{f} for frequency, \emph{a} for amplitude, \emph{cf} for color frequency and \emph{ca} for color amplitude. 

\bibliographystyle{abbrv-doi}
\bibliography{references.bib}